\documentclass[a4paper, 12pt]{article}
\usepackage{latexsym,amsmath,amsfonts,amssymb}
\usepackage{tikz}
\usetikzlibrary{decorations.pathmorphing,cd,decorations.markings,calc}
\usepackage{mathrsfs}
\usepackage[american]{babel}
\usepackage{graphicx}
\usepackage{bbm}
\usepackage{cite}
\usepackage{tcolorbox}
\usepackage{enumerate}
\usepackage{bbding}
\usepackage{colortbl}
\usepackage{xcolor}
\usepackage[font={small,it}]{caption}
\usepackage[normalem]{ulem}

\usepackage[colorlinks=true, citecolor=blue!90!black, linkcolor=blue!90!black, linktocpage=true, urlcolor=red!70!black]{hyperref}

\renewcommand{\baselinestretch}{1.05}

\usepackage[a4paper,
  left=2cm,
  right=2cm,
  top=2cm,
  bottom=2cm
]{geometry}

\newcommand{\ds}{\displaystyle}

\newcommand{\fP}{\mathfrak{P}}

\newcommand{\rev}[1]{{#1}}
\newcommand{\soutn}[1]{}

\numberwithin{equation}{section}

\newcommand{\be}{\begin{equation}} \newcommand{\ee}{\end{equation}}
\newcommand{\bea}{\begin{equation} \begin{aligned}} \newcommand{\eea}{\end{aligned} \end{equation}}

\newcommand{\cA}{\mathcal{A}}

\newcommand{\cB}{\mathcal{B}}
\newcommand{\cC}{\mathcal{C}}

\newcommand{\cH}{\mathcal{H}}

\newcommand{\cL}{\mathcal{L}}
\newcommand{\calL}{\mathscr{L}}
\newcommand{\cM}{\mathcal{M}}
\newcommand{\calM}{\mathscr{M}}

\newcommand{\calD}{\mathscr{D}}
\newcommand{\cN}{\mathcal{N}}
\newcommand{\cO}{\mathcal{O}}

\newcommand{\cS}{\mathcal{S}}
\newcommand{\cT}{\mathcal{T}}
\newcommand{\cU}{\mathcal{U}}
\newcommand{\cV}{\mathcal{V}}

\newcommand{\cZ}{\mathcal{Z}}

\newcommand{\bN}{\mathbb{N}}

\newcommand{\bR}{\mathbb{R}}

\newcommand{\bZ}{\mathbb{Z}}

\newcommand{\unit}{\mathbbm{1}}

\def\repa{\raise4pt\hbox{$\square$}\mkern-14mu\raise-4pt\hbox{$\square$}}
\def\repab{\overline{\raise4pt\hbox{$\square$}\mkern-14mu\raise-4pt\hbox{$\square$}\mkern-1mu}}

\newcommand{\abA}{\mathbb{A}}
\newcommand{\abB}{\mathbb{B}}

\begin{document}

\thispagestyle{empty}
\fontsize{12pt}{20pt}
\vspace{13mm}
\begin{center}
	{\huge Defect Charges, Gapped Boundary Conditions, \\[0.5em]
and the Symmetry TFT}
	\\[13mm]
    	{\large Christian Copetti$^a$}
	
	\bigskip
	{\it
		$^a$ Mathematical Institute, University of Oxford, Woodstock Road, Oxford, 
  
  OX2 6GG, United Kingdom.
	}
\end{center}

\bigskip

\begin{abstract}We offer a streamlined and computationally powerful characterization of higher representations (higher charges) for defect operators under generalized symmetries, employing the powerful framework of Symmetry TFT $\cZ(\cC)$. For a defect $\calD$ of codimension $p$, these representations (charges) are in one-to-one correspondence with gapped boundary conditions for the SymTFT $\cZ(\cC)$ on a manifold $Y = \Sigma_{d-p+1} \times S^{p-1}$, and can be efficiently described through dimensional reduction. We explore numerous applications of our construction, including scenarios where an anomalous bulk theory can host a symmetric defect. This generalizes the connection between 't Hooft anomalies and the absence of symmetric boundary conditions to defects of any codimension. Finally we describe some properties of surface charges for $3+1$d duality symmetries, which should be relevant to the study of Gukov-Witten operators in gauge theories.
\end{abstract}

\vspace{1.25cm}

\pagenumbering{arabic}
\setcounter{page}{1}
\setcounter{footnote}{0}
\renewcommand{\thefootnote}{\arabic{footnote}}

{\renewcommand{\baselinestretch}{.88} \parskip=0pt
\setcounter{tocdepth}{2}

\newpage

\tableofcontents}

\newpage

\section{Introduction}
\begin{figure}
    \centering
  \begin{tikzpicture}[scale=0.75]
 \draw[color=white!70!gray, fill=white, opacity=0.75] (0,0) -- (2,1) -- (2,3) -- (0,2) -- cycle;   \node at (1,1.5) {$X$};
\node[right] at (2.25,1.5) {$=$};
\begin{scope}[shift={(6.5,0)}]
     \draw[color=red, fill=white!90!red, opacity=0.75] (-3,0) -- (-1,1) -- (-1,3) -- (-3,2) -- cycle; \node[red] at (-2,1.5) {$\calL_{sym}$};
   \draw[white!75!black] (0,0) -- (-3,0);    \draw[white!75!black] (2,1) -- (-1,1);   \draw[white!75!black] (2,3) -- (-1,3);   \draw[white!75!black] (0,2) -- (-3,2);        \draw[color=white!70!blue, fill=white!90!gray, opacity=0.75] (0,0) -- (2,1) -- (2,3) -- (0,2) -- cycle;   \node[blue] at (1,1.5) {$X_{rel}$};
\end{scope}
\end{tikzpicture} \hspace{3cm}
\begin{tikzpicture}[scale=0.75]
     \draw[color=white!70!blue, fill=white, opacity=0.75] (0,0) -- (2,1) -- (2,3) -- (0,2) -- cycle;   
     \draw[red,thick] (1,1.5) -- (1,2.5) node[above] {$\cL$};
     \draw[fill=black] (1,1.5) node[below] {$\lambda$} circle (0.05);
     
\node[right] at (2.25,1.5) {$=$};
\begin{scope}[shift={(6.5,0)}]
     \draw[color=red, fill=white!90!red, opacity=0.75] (-3,0) -- (-1,1) -- (-1,3) -- (-3,2) -- cycle;
     \draw[red,thick] (-2,1.5) -- (-2,2.5) node[above] {$\cL$};
          \draw[fill=red] (-2,1.5) circle (0.05);
     \draw (-2,1.5) -- (1,1.5); \node[above] at (-0.5,1.5) {$\lambda$};
              \draw[fill=black] (1,1.5) circle (0.05);
   \draw[white!75!black] (0,0) -- (-3,0);    \draw[white!75!black] (2,1) -- (-1,1);   \draw[white!75!black] (2,3) -- (-1,3);   \draw[white!75!black] (0,2) -- (-3,2);        \draw[color=white!70!gray, fill=white!90!gray, opacity=0.75] (0,0) -- (2,1) -- (2,3) -- (0,2) -- cycle;  
\end{scope}
\end{tikzpicture}
    \caption{SymTFT setup. Left the sandwich construction for the theory $X$, right the identification of charged multiplets.}
    \label{fig: symtft}
\end{figure}
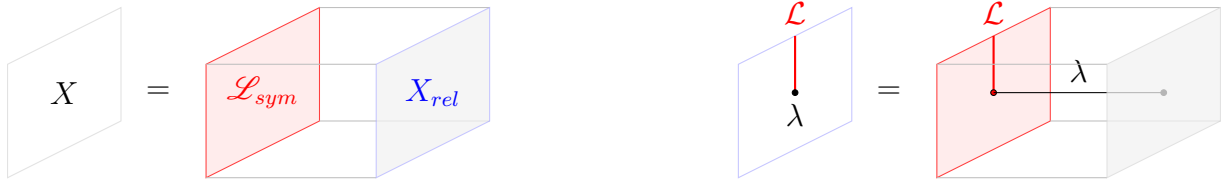
Generalised symmetries \cite{Gaiotto:2014kfa} provide an elegant tool to deepen our understanding of strongly coupled physical systems. A key aspect of their power derives from how these symmetries act on charged operators. Such action is typically implemented via linking, as discussed in \cite{Gaiotto:2014kfa} and many subsequent works. However, this is not the sole form of symmetry action. Bulk topological defects may or may not terminate in a topologically consistent manner on the charged object. We refer to this collection of data as a \emph{defect charge} or \emph{defect multiplet}. Such action is relevant when discussing extended charged objects, such as boundaries, interfaces and extended defects of higher codimension.\footnote{The study of defects, especially conformal one, has been a very fruitful one so far. See e.g. \cite{Liendo:2012hy,Billo:2016cpy,Herzog:2017xha} for some classic references on the subject.}

Clearly understanding and characterising the structure of these multiplets is crucial for addressing the kinematical constraints of symmetry. This note provides a unified description of multiplets through dimensionally reduced gapped boundary conditions in the Symmetry TFT, presenting a clear and concrete framework.

We hope that these results can be applied to the description of defect RG flows, for example by constraining the structure of IR defect multiplets and the permissible transitions induced by defect deformations.

\subsection{(Higher) Charges and the SymTFT}
Given a symmetry category $\cC$\footnote{For recent reviews on generalized symmetries see e.g. \cite{Shao:2023gho,Schafer-Nameki:2023jdn}.} a natural question is what are its allowed representations/multiplets. Mathematically a ``representation" is encoded in the correct notion of (higher) Module Category over $\cC$. However, this soon becomes a daunting description and a more direct computational tool would be welcome.\footnote{For 1-Categories, module categories are textbook material \cite{etingof2016tensor}, For higher categories, while in principle clear, has not been flashed out in full generality. See however \cite{schommer2009classification,Decoppet:2021skp,Bhardwaj:2022maz} for material concerning Module 2-Categories.} 

A complementary viewpoint is provided by the SymTFT picture \cite{Freed:2022qnc,Apruzzi:2021nmk,Ji:2021esj,Chatterjee:2022kxb}, which identifies a QFT $X$ with symmetry $\cC$ with the interval compactification of a triplet:
\be
\left( \, \cZ(\cC), \, \calL_{sym}, \, X_{\text{phys}} \, \right) \, ,
\ee
where $\cZ(\cC)$ is the Drinfeld center of $\cC$ and its objects describe a $d+1$ dimensional TFT which we denote by the same name; $\calL_{sym}$ is a canonical Dirichlet gapped boundary condition for $\cZ(\cC)$, which hosts on its worldvolume topological defects $\cL$ belonging to the symmetry category $\cC$ and $X_{phys}$ is a free, dynamical boundary condition which couples the dynamics of $X$ to its symmetry. The above information is usually compressed into a sandwich picture, see Figure \rev{\ref{fig: symtft}}.

The power of this construction is that, apart from the symmetry $\cC$, it also encodes its (higher) representations $\lambda$. These have been referred to in the existing literature as generalized charges \cite{Bhardwaj:2023wzd,Bhardwaj:2023ayw} or higher Tube-algebra \cite{Bartsch:2023pzl,Bartsch:2023wvv}.\footnote{Since there is no consensus about which denomination to use we will use the terms generalized/higher charges/representations/multiplets interchangeably.} 
A generalized charge $(\lambda, \cL)$ is encoded in an object $\lambda$ of the SymTFT connecting the $X_{phys}$ boundary to a symmetry defect $\cL$ of the same codimension on $\calL_{sym}$. This describes a charged multiplet residing in the $\cL$-twisted Hilbert space, see Figure \ref{fig: symtft}.

The characterization of the complete set $\lambda$ of objects is far from obvious in the bulk description. For instance, it may also include condensation defects \cite{Roumpedakis:2022aik}. This description formed part of the original SymTFT proposal for charged local operators and was extended by \cite{Bhardwaj:2023ayw} and \cite{Bartsch:2023wvv} to encompass a class of extended defects. The purpose of this note is to offer an alternative characterization through the lens of gapped boundary conditions, along with some intriguing physical insights.

We believe this endeavor to be worthwhile, as current methods for addressing such questions often depend heavily on categorical machinery. In higher-dimensional cases, this machinery can be extremely abstract or not yet fully developed. Thus, developing a concrete computational tool holds clear physical interest. Additionally, our approach will standardize the construction of all types of multiplets, providing a unified perspective on them. Finally, our methods readily reveal the internal structure of defect operator charges, i.e., charges \emph{within} the defect itself, which have not been extensively discussed in the generalized symmetries literature.

While this Note is mostly of technical nature, we hope to report soon on its interesting physical applications.

This work leverages heavily on results developed in \cite{CCboundaries} to describe boundary conditions in the SymTFT framework. After the present work appeared we learned that several other groups \cite{Cordova:2024vsq,Choi:2024tri} were pursuing similar ideas. 

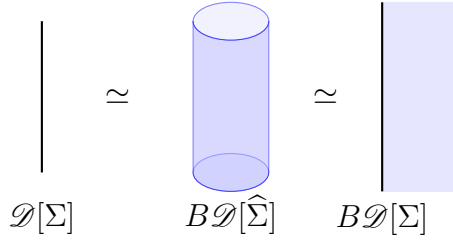
\begin{figure}
    \centering
   \begin{tikzpicture}
 \draw[black, thick] (0,0) -- (0,2);   
 \node[below] at (0,-0.25) {$\calD[\Sigma]$};
 \node at (1,1) {$\simeq$};
 \begin{scope}[shift={(2.5,0)}]
 \draw[color=blue, fill=white!90!blue, opacity=0.5] (0,0) ellipse (0.5 and 0.25);    
  \draw[color=blue, fill=white!70!blue, opacity=0.5] (-0.5,0) -- (-0.5,2) arc (-180:0:0.5 and 0.25) -- (0.5,0) arc (0:-180:0.5 and 0.25) -- cycle;
  \draw[color=blue, fill=white!90!blue, opacity=0.5] (0,2) ellipse (0.5 and 0.25);    
  \node at (1.25,1) {$\simeq$};
  \node[below] at (0,-0.15) {$B\calD[\widehat{\Sigma}]$};
 \end{scope}
 \begin{scope}[shift={(4.5,0)}]
 \draw[color=white!90!blue, fill=white!90!blue] (0,-0.25) -- (1,-0.25) -- (1,2.25) -- (0,2.25) --cycle;
 \draw[thick,black] (0,-0.25) node[below] {$B\calD[\Sigma]$} -- (0,2.25);
 \end{scope}
\end{tikzpicture}
    \caption{Correspondence between defects and boundary conditions. First we excise a neighbourhood bounded by $\widehat{\Sigma}$ from spacetime to obtain a boundary condition $B\calD$. Finally, reducing on the sphere $S^{p-1}$, we study a related boundary condition in the dimensionally reduced bulk theory.}
    \label{fig: defectcorr}
\end{figure}

\subsection{Higher representations and boundary conditions} 
Let us outline our prescription, which we will describe in detail in the next Section \ref{sec: reps}. Recall that extended defects in QFT often have a ``disorder"-type definition as follows. For a codimension $p$ defect $\calD$ with worldvolume $\Sigma$, we excise from spacetime a cylindrical region with boundary $\widehat{\Sigma} = \Sigma \times S^{p-1}_\epsilon$, where $S^{p-1}_\epsilon$ is a $(p-1)$-dimensional sphere of radius $\epsilon$ centered around the worldvolume $\Sigma$ of the defect. The parameter $\epsilon$ is a UV regulator, chosen to be smaller than any physical scale in the theory. This defines a boundary condition $B\calD$ on $\widehat{\Sigma}$ corresponding to a defect of type $\calD$.\footnote{For $p=1$, we take $S^0$ to be a disjoint union of two points. Following the steps below, one recovers the well-known fact that interfaces in the theory $X$ are described by boundary conditions in the folded theory $X \boxtimes \overline{X}$. In this note and will focus mainly on higher codimensional defects.}

If the bulk-defect system is conformal, this can be made precise by mapping $\widehat{\Sigma}$ to the conformal boundary of $\text{AdS}_{d-p + 1} \times S^{p-1}$, as pioneered by Kapustin \cite{Kapustin:2005py}. In this case $\epsilon$ is identified with the \soutn{standard} radial cutoff in AdS. Performing a KK reduction on $S^{p-1}$ connects this to a boundary condition on $\Sigma$ for a $d-p+1$-dimensional theory. The general setup is presented in Figure \ref{fig: defectcorr}

While it is not obvious whether \emph{order} operators can also be given a similar definition, it is expected, at least in the conformal setup, that a sort of state-defect correspondence should continue to hold, although with the needed precautions. See \cite{Hofman:2024oze} for a recent study.
Nevertheless, besides the obvious complications that arise if conformality is forsaken, as long as we are interested in the  $\cC$-symmetry action on $\calD$ \emph{only}, it is only the topology of $\widehat{\Sigma}$ that matters.

Once this is established, there is a natural guess for the SymTFT description of the higher representations-charges. A boundary condition $B_a$ -- $a$ being the label on which the symmetry representation acts --  in the SymTFT corresponds to the choice of a second gapped boundary $\calL_B$ stretching between the physical boundary and the ``symmetry" topological boundary $\calL_{sym}$. Their intersection is labelled by an element $a$ of the $\cC$ module category corresponding to $\calL_B$ \cite{Huang:2023pyk,CCboundaries}.\footnote{Indeed it is known that module categories $\cM$ over $\cC$ are in correspondence with the Lagrangian algebras $\calL'$ in the Drinfeld center $\cZ(\cC)$. This is a theorem for symmetries in $1+1$ dimensional systems and solid folklore in higher dimensions.}

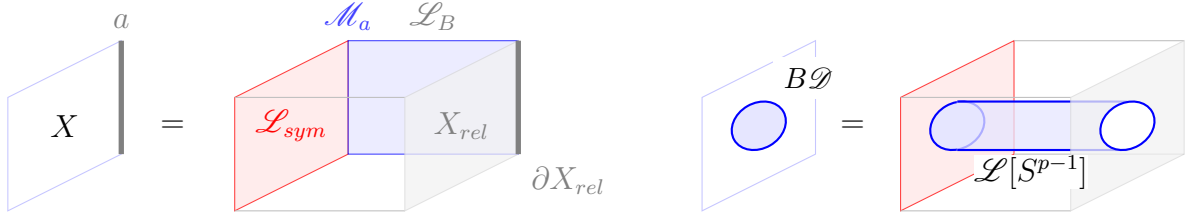
\begin{figure}
    \centering
   \begin{tikzpicture}[scale=0.75]
    \draw[color=white!70!blue, fill=white, opacity=0.75] (0,0) -- (2,1) -- (2,3) -- (0,2) -- cycle;   
\draw[gray, line width =2] (2,1) -- (2,3) node[above] {$a$};
\node at (1,1.5) {$X$};
\node[right] at (2.5,1.5) {$=$};
\begin{scope}[shift={(7,0)}]
     \draw[color=red, fill=white!90!red, opacity=0.75] (-3,0) -- (-1,1) -- (-1,3) -- (-3,2) -- cycle;
   \node[red] at (-2,1.5) {$\calL_{sym}$};
     \draw[color=blue, fill=white!90!blue, opacity=0.75] (-1,1) -- (-1,3) node[above] {$\calM_a$} -- (2,3) -- (2,1) -- cycle;
     \draw[white!75!black] (0,0) -- (-3,0); \draw[white!75!black] (0,2) -- (-3,2); 
     \node[gray,above] at (0.5,3) {$\calL_{B}$};
      \draw[color=white!70!gray, fill=white!90!gray, opacity=0.75] (0,0) -- (2,1) -- (2,3) -- (0,2) -- cycle;
      \node[gray] at (1,1.5) {$X_{rel}$};
      \draw[gray, line width =2] (2,1)  node[below right,gray] {$\partial X_{rel}$} -- (2,3);
      \end{scope}
\end{tikzpicture}  \hspace{1cm}\begin{tikzpicture}[scale=0.75]
  \draw[color=white!70!blue, fill=white, opacity=0.75] (0,0) -- (2,1) -- (2,3) -- (0,2) -- cycle; 

 \draw[rotate around={30:(1,1.5)},thick, blue, fill=white!90!blue] (1,1.5) ellipse (0.5 and 0.4);
 \node[right, fill=white] at (1.25,2.35) {\small$B\calD$};
\node[right] at (2.25,1.5) {$=$};
\begin{scope}[shift={(6.5,0)}]
     \draw[color=red, fill=white!90!red, opacity=0.75] (-3,0) -- (-1,1) -- (-1,3) -- (-3,2) -- cycle;
      \draw[rotate around={30:(-2,1.5)},thick, blue, fill=white!90!blue] (-2,1.5) ellipse (0.5 and 0.4);

\pgfmathsetmacro{\yA}{1.5 + sqrt((0.4*cos(30))^2 + (0.5*sin(30))^2 ) };
\pgfmathsetmacro{\yB}{1.5 - sqrt((0.4*cos(30))^2 + (0.5*sin(30))^2 ) };

\draw[color=white!90!blue, fill=white!90!blue,opacity=0.75] (-2,\yA) -- (1,\yA)-- (1,\yB)  -- (-2,\yB) -- cycle;
\draw[blue,thick] (-2,\yA) -- (1,\yA); \draw[blue,thick] (-2,\yB) -- (1,\yB);

   \draw[white!75!black] (0,0) -- (-3,0);    \draw[white!75!black] (2,1) -- (-1,1);   \draw[white!75!black] (2,3) -- (-1,3);   \draw[white!75!black] (0,2) -- (-3,2);        \draw[color=white!70!gray, fill=white!90!gray, opacity=0.75] (0,0) -- (2,1) -- (2,3) -- (0,2) -- cycle;   
         \draw[rotate around={30:(1,1.5)},thick, blue, fill=white] (1,1.5) ellipse (0.5 and 0.4);
         \node[above, inner sep=0pt, fill=white, opacity=1] at ($(-0.5,\yA) + (-0.2,-1.55)$) {$\calL[S^{p-1}]$};
\end{scope}
\end{tikzpicture}
    \caption{Sym TFT setup for a boundary condition (Left) an for a defect (Right).}
    \label{fig: symtftbdydef}
\end{figure}

Similarly, the defect boundary condition $B\calD$ on $\widehat{\Sigma}$ extends into the bulk to a gapped boundary condition 
\be \calL[S^{p-1}]: \ \ \text{defined on} \ \ \widehat{\Sigma} \times I = \underbrace{\left( \Sigma \times I \right)}_{Y_{d-p+1}} \times S^{p-1} 
\ee
ending on the symmetry b.c. $\calL_{sym}$. The setup is shown in Figure \ref{fig: symtftbdydef}.

This can be thought of as a ``magnetic" description of the bulk SymTFT defects.\footnote{We thank Andrea Antinucci for discussion on this point.}
We will henceforth use the notation $P[\Sigma]$ to denote the dimensional reduction of an object $P$ on the compact manifold $\Sigma$. 
Crucially, since the topology around the defect $\calD$ is fixed, the problem of understanding the its (higher) charges boils down to the description of gapped boundary conditions $\calL[S^{p-1}]$ on a fixed topology $Y_{d-p+1} \times S^{p-1}$. These form a (very) different set from universal boundary conditions, which are required to exist on any codimension-one manifold. Said otherwise, a gapped boundary condition $\calL[S^{p-1}]$ \emph{does not} necessarily descend from the dimensional reduction of a full fledged gapped b.c. $\calL$.
\footnote{A well known related example is the SymTFT realization of class $\cS$ theories \cite{Bashmakov:2022uek,Antinucci:2022cdi}. In this case the bulk $7d$ CS theory admits no gapped boundary conditions on general $Y_6$, however there are various consistent choices once the Gaiotto curve $\Sigma_g$ is fixed and $Y_6 = Y_4 \times \Sigma_g$, which makes class $\cS$ theories into absolute QFTs, contrary to their $6d$ $\cN=(2,0)$ counterpart.} Thus we arrive at our first punchline:

\begin{center}
    \textit{A ``defect charge"} $\calL[S^{p-1}]$ \textit{of codimension} $p$ \textit{is described by a gapped boundary condition for the dimensionally reduced SymTFT} $\cZ(\cC)[S^{p-1}]$.
\end{center}
\rev{In most of the following discussion, we will assume that no local topological operators arise in the compactification $\cZ(\cC)[S^{p-1}]$. When this assumption fails, the prescriptions below will need to be supplemented by an appropriate projection on the correct ``universe". This is related to the fact that the reduced SymTFT $\cZ(\cC)[S^{p-1}]$ might not have a simple unit (i.e. the reduced theory is described by a multi-fusion category). We will discuss this at the relevant steps.
Furthermore, in this paper we will mostly be concerned higher-codimensional defects $(p>1)$. For $p=1$, which describes interfaces, we adopt the convention that $S^0 = \{pt\} \cup \{pt\}$ is the disjoint union of two points. The ensuing compactification corresponds to the folding trick, under which $\cZ(\cC)[S^0] = \cZ(\cC\boxtimes\overline{\cC})$ with $\calL_{sym}[S^0] = \calL_{sym} \otimes \overline{\calL_{sym}}$. This again reduces to the study of boundary conditions, which are treated in detail in \cite{CCboundaries}.}

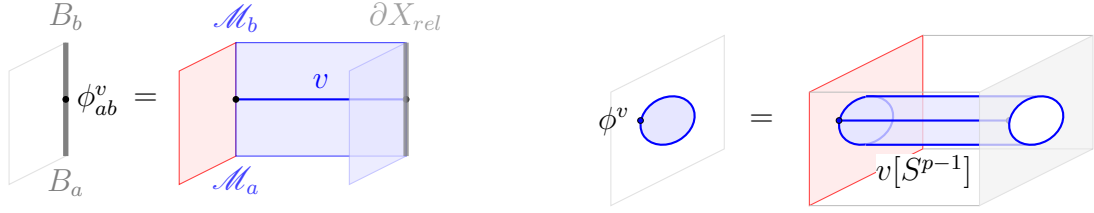
\begin{figure}[t]
    \centering
    \begin{tikzpicture}[scale=0.75]
    \draw[color=white!70!gray, fill=white, opacity=0.75] (1,0.5) -- (2,1) -- (2,3) -- (1,2.5) -- cycle;   
\draw[gray, line width =2] (2,1) node[below] {$B_a$} -- (2,3) node[above] {$B_b$};
\draw[fill=black] (2,2) node[right] {$\phi^v_{a b}$} circle (0.05);
\node[right] at (3,2) {$=$};
\begin{scope}[shift={(6,0)}]
     \draw[color=red, fill=white!90!red, opacity=0.75] (-2,0.5) -- (-1,1) -- (-1,3) -- (-2,2.5) -- cycle;
     \draw[color=blue, fill=white!90!blue, opacity=0.75] (-1,1) node[below] {$\calM_a$} -- (-1,3) node[above] {$\calM_b$} -- (2,3) -- (2,1) -- cycle;
     \draw[blue,thick] (-1,2) -- (2,2); \node[blue,above] at (0.5,2) {$v$}; \draw[fill=black] (-1,2) circle (0.05); \draw[fill=black] (2,2) circle (0.05);
      \draw[color=white!70!blue, fill=white!90!blue, opacity=0.75] (1,0.5) -- (2,1) -- (2,3) -- (1,2.5) -- cycle;
      \draw[gray, line width =2,opacity=0.75] (2,1) -- (2,3) node[above,gray] {$\partial X_{rel}$};
      \end{scope}
\end{tikzpicture} \hspace{1.5cm} \begin{tikzpicture}[scale=0.75]
  \draw[color=white!70!gray, fill=white, opacity=0.75] (0,0) -- (2,1) -- (2,3) -- (0,2) -- cycle; 
 \draw[rotate around={30:(1,1.5)},thick, blue, fill=white!90!blue] (1,1.5) ellipse (0.5 and 0.4);

\pgfmathsetmacro{\xA}{1 - sqrt((0.4*sin(30))^2 + (0.5*cos(30))^2 ) };
\pgfmathsetmacro{\xB}{-2 - sqrt((0.4*sin(30))^2 + (0.5*cos(30))^2 ) };
 \draw[fill=blue] (\xA, 1.5) node[left] {$\phi^v$} circle (0.05);

\node[right] at (2.25,1.5) {$=$};
\begin{scope}[shift={(6.5,0)}]
     \draw[color=red, fill=white!90!red, opacity=0.75] (-3,0) -- (-1,1) -- (-1,3) -- (-3,2) -- cycle;
      \draw[rotate around={30:(-2,1.5)},thick, blue, fill=white!90!blue] (-2,1.5) ellipse (0.5 and 0.4);

\pgfmathsetmacro{\yA}{1.5 + sqrt((0.4*cos(30))^2 + (0.5*sin(30))^2 ) };
\pgfmathsetmacro{\yB}{1.5 - sqrt((0.4*cos(30))^2 + (0.5*sin(30))^2 ) };
\pgfmathsetmacro{\xA}{1 - sqrt((0.4*sin(30))^2 + (0.5*cos(30))^2 ) };
\pgfmathsetmacro{\xB}{-2 - sqrt((0.4*sin(30))^2 + (0.5*cos(30))^2 ) };

\draw[color=white!90!blue, fill=white!90!blue, opacity=0.75] (-2,\yA) -- (1,\yA)-- (1,\yB)  -- (-2,\yB) -- cycle;
\draw[blue,thick] (-2,\yA) -- (1,\yA); \draw[blue,thick] (-2,\yB) -- (1,\yB);

\draw[thick,color=blue] (\xA,1.5) -- (\xB,1.5);
 \draw[fill=blue] (\xA, 1.5)  circle (0.05);
  \draw[fill=blue] (\xB, 1.5)  circle (0.05);

   \draw[white!75!black] (0,0) -- (-3,0);    \draw[white!75!black] (2,1) -- (-1,1);   \draw[white!75!black] (2,3) -- (-1,3);   \draw[white!75!black] (0,2) -- (-3,2);    
  
   \draw[color=white!70!gray, fill=white!90!gray, opacity=0.75] (0,0) -- (2,1) -- (2,3) -- (0,2) -- cycle;   
         \draw[rotate around={30:(1,1.5)},thick, blue, fill=white] (1,1.5) ellipse (0.5 and 0.4);
            \node[above, inner sep=0pt, fill=white, opacity=1] at ($(\xB,1.5) + (1.5,-1.2)$) {$v[S^{p-1}]$};
\end{scope}
\end{tikzpicture}
    \caption{SymTFT setup for a boundary multiplet (Right) and for a defect multiplet (Left).}
    \label{fig: symtftbdycharge}
\end{figure}

\subsection{\soutn{Multiplets within multiplets} \rev{Defect Operator Multiplets}}
Another notable feature of this approach is the ability to describe in an intuitive manner charged excitations $v$ on a defect $\calD$. These include defect-changing interfaces (of codimension one on the defect worldvolume) as well as defect operators of various dimensionalities. We will call these defect \emph{operator} multiplets to avoid confusion.

Again it is useful to review the case of a boundary condition first \cite{CCboundaries,Copetti:2024dcz}. Given a boundary condition $B_a$ and the associated topological boundary $\calL_B$, allowed boundary multiplets $\phi^v_{a b}$ are described by topological defects $v$ confined to the $\calL_B$ boundary and stretching between $\calL_{sym}$ and $X_{rel}$. See Figure \ref{fig: symtftbdycharge}.

Clearly we can have $a \neq b$ only if $v$ is of codimension one on $\calL_B$. This description has a number of interesting applications, from the description of boundary changing operators in CFT \cite{CCboundaries} (see also \cite{Fuchs:2002cm} for an equivalent characterization when $\cC$ is a braided category) to that of massive kinks \cite{Copetti:2024dcz,Copetti:2024rqj,Cordova:2024vsq}. A related mathematical description also appears in the context of anyon chain models \cite{Aasen:2020jwb,Buican:2017rxc,Bhardwaj:2024kvy}.
The generalization of this prescription to defects is straightforward, one simply considers objects $v[S^{p-1}]$ confined on the reduced boundary condition $\calL[S^{p-1}]$, see Figure \ref{fig: symtftbdycharge}.
We thus arrive at the following prescription:
\begin{center}
    \textit{Defect operator multiplets} $v$ \textit{are described by topological operators in} $\calL[S^{p-1}]$ \textit{ending on the intersection $\calM$ between} $\calL_{sym}$ \textit{and $\calL[S^{p-1}]$}.
\end{center}

The program of understanding higher charges then consists of three steps, similar to the standard SymTFT picture:
\begin{enumerate}[i)]
    \item Classify gapped boundary conditions $\calL[S^{p-1}]$ for $\cZ(\cC)$ on $Y_{d-p + 1} \times S^{p-1}$. This classification can be studied by describing the dimensionally reduced SymTFT on $S^{p-1}$. This setup has already appeared in \cite{Nardoni:2024sos} to describe the symmetries of dimensionally reduced QFTs.
    \item Describe the topological junction $\calM_{\calL[S^{p-1}], \calL_{sym}[S^{p-1}]}$ with the symmetry boundary condition $\calL_{sym}$. These describe the module category structure for genuine defect operators. A similar problem can also be considered for twisted defects, though we do not explore this in full generality in the present note.
    \item Describe the symmetry action on defect charges and defect operator multiplets.
\end{enumerate}

\begin{figure}
    \centering
    \begin{tikzpicture}[scale=0.75]
      \draw[color=red, fill=white!90!red, opacity=0.75] (-3,0) -- (-1,1) -- (-1,3) -- (-3,2) -- cycle;
   \node[red, above right] at (-3,0.55) {$\calL_{sym}$};
     \draw[color=blue, fill=white!90!blue, opacity=0.75] (-1,1) -- (-1,3) node[above] {$\calM_a$} -- (2,3) -- (2,1) -- cycle;
     \node[blue,above] at (0.5,3) {$\calL_{B}$};
      \draw[white!75!black] (0,0) -- (-3,0); \draw[white!75!black] (0,2) -- (-3,2);
     \draw[->] plot[smooth, tension=1.5] coordinates{(-1.5,2) (-1,1.9) (-0.5,2.25)};
       \draw[->] plot[smooth, tension=1.5] coordinates{(-1.5,1.5) (-1,1.4) (-0.5,1.75)};
       \draw[color=white!70!blue, fill=white!90!blue, opacity=0.75] (0,0) -- (2,1) -- (2,3) -- (0,2) -- cycle;
      \node[gray] at (1,1.5) {$X_{rel}$};
        \draw[gray, line width =2] (2,1)  -- (2,3);
       \node at (3,1.5) {$=$};
       \begin{scope}[shift={(4,0)}]
       \draw[color=white!70!blue, opacity=0.75] (0,0) -- (2,1) -- (2,3) -- (0,2) -- cycle;  
       \draw[color=white!70!blue, fill=white!90!blue, opacity=0.75] (1,0.5) -- (2,1) -- (2,3) -- (1,2.5) --cycle;
\draw[gray, line width =2] (2,1) -- (2,3) node[above] {$\calM_a$};
\node at (1.5,1.75) {$\cT_B$};    
\node at (0.5,1.25) {$X$};    
\draw[gray, line width =2, dashed] (1,0.5) -- (1,2.5);
       \end{scope}
    \end{tikzpicture}
    \caption{Wedge compactification allows to describe a boundary condition as a transparant interface between $X$ and a gapped theory $\cT_B$.}
    \label{fig: wedgecomp}
\end{figure}
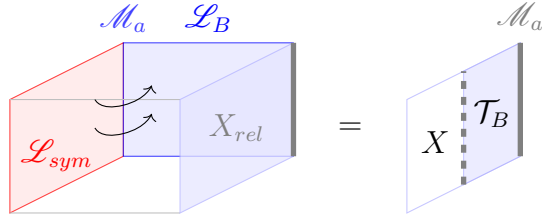

\subsection{Multiplets and TFTs}
In \cite{CCboundaries}, it is shown that a boundary condition $B_a$ can be viewed as a $\cC$-transparent interface between the theory $X$ and a TFT $\cT_B$ with boundary condition $a$. This follows from the observation that the wedge compactification between $\calL_{sym}$ and $\calL_B$ describes a $\cC$-symmetric gapped phase $\cT_B$, as discussed in \cite{Bhardwaj:2023idu,Bhardwaj:2023fca,Huang:2023pyk,Chatterjee:2022tyg} (see Figure \ref{fig: wedgecomp}).

The theorem from \cite{Jensen:2017eof,Thorngren:2020yht} (\rev{which can be extended to non-invertible symmetries in 1+1d by the results of \cite{Choi:2023xjw}),} states that anomalous symmetries do not admit \emph{\rev{strongly} symmetric} boundary conditions, can then be given a straightforward proof.\footnote{Inspired by \cite{Choi:2023xjw} we define a \rev{strongly} symmetric defect to be a defect \soutn{ which all objects in $\cC$ can end topologically.} \rev{which is invariant under parallel fusion with the bulk topological defects.} More about this will be explained in Section \ref{sec: reps}.} Specifically, the presence of a 't Hooft anomaly prohibits a trivial symmetric gapped phase and such a phase would result in a \rev{strongly} symmetric interface upon the SymTFT construction.\footnote{This also raises an interesting puzzle, since it is not always possible to saturate 't Hooft anomalies through gapped phases, as exemplified by the cubic $U(1)$ anomaly or local gravitational anomalies \cite{Cordova:2019bsd}. Understanding how the SymTFT can describe such instances would be valuable. The author thanks K. Ohmori for highlighting this point.}

A similar conclusion can be extended to defects by compactifying the SymTFT on $S^{p-1}$. The symmetry action on the defect is equivalent to a symmetric interface between the defect world-volume and a gapped phase $\cT_B[S^{p-1}]$ for the dimensionally reduced symmetry $\cC[S^{p-1}]$, with boundary condition $a$. We comment upon the rich landscape associated to such a picture in Section \ref{sec: anom}.

\begin{center}\rule{0.15\textwidth}{0.1pt}\end{center}
\vspace{0.75cm}

The plan for the rest of the note is as follows. In
\textbf{Section \ref{sec: reps}} we introduce the relevant notation and study the problem of defining boundary conditions for the compactified theory, in \textbf{Section \ref{sec: examples}} we give several simple examples of explicit computations.
   In \textbf{Section \ref{sec: examplesduality}} we focus on the multiplet structure for duality defects and its interpretation in the context of GW operators. 
    In \textbf{ Section \ref{sec: anom}} we discuss the constraints imposed by 't Hooft anomalies on defect multiplets, extending \cite{Thorngren:2020yht}.
   We conclude with a brief discussion of open research directions.

\section{Symmetry, multiplets, and gapped boundaries}\label{sec: reps}
In this Section we will give more details about symmetry action on defects and its SymTFT description. Since the mathematical framework surrounding these ideas is not completely developed, some parts will not try to be comprehensive. We however try to amend this in later Sections when we present various explicit examples.

\subsection{Symmetry action in QFT}In order to motivate the discussion, let us review two well-known ways in which symmetry can be implemented on objects in QFT: linking  and topological junctions. This will allow us to justify what we mean by saying that a defect is \emph{symmetric} or, on the other end of the spectrum, that it \emph{\soutn{spontaneously} breaks} the symmetry. A similar discussion for boundary conditions is beautifully outlined in \cite{Choi:2023xjw}.

\vspace{2mm}\noindent\textbf{Linking} This type of action is well known since \cite{Gaiotto:2014kfa}. Given a codimension $p$ defect $\calD$, this can be charged under a $d-p$-form symmetry $G^{(d-p)}$.\footnote{We assume that $G$ is Abelian also when $d-p=0$ for simplicity.} The charge is defined by linking the symmetry generator $U$ with defect through the transverse $S^{p-1}$:
\bea
\begin{tikzpicture}[baseline={(0,1)}]
  \draw[thick] (0,0) node[below] {$\calD$} -- (0,0.7); \draw[thick] (0,0.8) -- (0,2);
  \draw[red,thick] (0, 0.75) arc(-90:-260: 0.5 and 0.25);  \draw[red,thick] (0, 0.75) arc(-90:80: 0.5 and 0.25);
  \node[left,red] at (-0.5,1) {$U$};
\end{tikzpicture} = u(\calD)  \begin{tikzpicture} [baseline={(0,1)}]
    \draw[thick] (0,0) node[below] {$\calD$} -- (0,2);
\end{tikzpicture}
\eea
$u \in G^\vee$ being a character. This has a simple interpretation from the dimensional reduction viewpoint: after reducing on $S^{p-1}$ the linking operator becomes a pointlike object $\phi \equiv U[S^{p-1}]$ and its vev on the defect is just the charge:
\be
\langle \phi \rangle_{\calD} = u(\calD) \, .
\ee

\vspace{2mm}\noindent \textbf{Topological junctions} On the other end of the spectrum we have the action of topological defects of the same (or smaller) codimension as $\calD$ by parallel fusion.
This is better understood by introducing topological domain walls between a topological defect $\cL$ and a dynamical defect $\calD$:
\bea
\begin{tikzpicture}
 \draw[thick] (0,0) node[below] {$\calD$} -- (0,2) node[above] {$\calD'$};
 \draw[thick,red] (-1,1) node[above] {$\cL$} -- (0,1);
 \draw[fill=red] (0,1) node[right,black] {$e_\cL^{\calD \, \calD'}$} circle (0.05);
\end{tikzpicture}
\eea
\soutn{we will say that a symmetry $\cL$ is \emph{preserved} by $\calD$ if all topological junctions $e_\cL$ leave the defect invariant $\calD'=\calD$. In codimension one, this corresponds to the notion of \emph{strongly symmetric} boundary condition \cite{Choi:2023xjw}.} 
\rev{We will say that a symmetry $\cL$ is \emph{preserved} by $\calD$ if topological junctions $e_\cL$ always leave the defect invariant $\calD'=\calD$. In codimension one, this corresponds to the notion of \emph{strongly symmetric} boundary condition \cite{Choi:2023xjw}. We will thus denote $\calD$ as a (strongly) \emph{symmetric} defect.}
Similarly, higher codimension topological defects $\cL^{(q)}$, with $q>p$ can end on $\calD$ topologically, forming a junction with a topological defect $\cL^{(q)}_\calD$ on $\calD$. In this case we will say that $\cL^{(p)}$ is \emph{preserved} by the defect if it can \soutn{only} end topologically on the trivial defect line $\unit^{(q)}_\calD$. 
\rev{If this cannot happen, the symmetry is broken by the defect.}\footnote{\rev{In certain applications, it might be more natural to define a symmetric defect in the weak sense \cite{Choi:2023xjw}, that is, by just requiring the existence of a topological junction for the symmetry $\cL$ on $\calD$. For non-invertible symmetries, this is a much weaker notion than invariance under fusion. This definition is perfectly acceptable, as the junction allows to define a symmetry action on the defect Hilbert space. In this paper, however, we will focus on the description of \emph{strong} symmetry breaking.}}
\soutn{Otherwise, we will say that the symmetry is \emph{spontaneously broken by the defect}.} 
A defect preserving the whole symmetry category $\cC$ \rev{as per our definition} is called \emph{$\cC$-symmetric}. We will give a SymTFT justification for this definition below.\footnote{Notice that this coincides with the standard definition for 0-form symmetries acting on boundary conditions, as we can describe a defect $\cL$ which cannot terminate topologically on $\calD$ as the fusion product $\cL \times \calD$, which is a codimension-0 defect for the $\calD$ multiplet.}
Topological junctions implement a \emph{defect symmetry} under which defect operators $v$ might be charged:
\bea \label{eq: defectsymm}
\begin{tikzpicture}
     \draw[thick] (0,0) node[below] {$\calD$} -- (0,2);
 \draw[thick,red] (-1,0.5) node[above] {$\cL$} -- (0,0.5);
 \draw[fill=red] (0,0.5) circle (0.05);
 \draw[fill=black] (0,1.5) node[right] {$v$} circle (0.05);
 \node[right] at (0.5,1) {$\ds = \cL_{\calD}[v]$};
 \begin{scope}[shift={(3.5,0)}]
         \draw[thick] (0,0) node[below] {$\calD$} -- (0,2);
 \draw[thick,red] (-1,1.5) node[above] {$\cL$} -- (0,1.5);
 \draw[fill=red] (0,1.5) circle (0.05);
 \draw[fill=black] (0,0.5) node[right] {$v$} circle (0.05);
 \end{scope}
    \end{tikzpicture}
\eea
They also feature nontrivial composition properties, which reflect the product structure on the bulk defects. Given two bulk defects $\cL, \, \cL'$ and topological junctions $e_\cL, \, e_{\cL'}$ the bulk fusion $\cL \times \cL' = \bigoplus_{\cL''} N_{\cL \cL'}^{\cL''} \, \cL''$ induces a defect junction $f_{\cL \cL'}^{\cL''}$ between $e_\cL \times e_{\cL'}$ and $e_{\cL''}$. 
This structure continues until we reach point-lke junctions, which are related to each other by linear maps.
For boundary conditions in $1+1$d the relevant mathematical structure is that of a $\cC$-module category and is nicely summarized in e.g. \cite{etingof2016tensor,Choi:2023xjw,Cordova:2024vsq}.

Finally, topological defects with $q<p$ can -- if $d-q \geq p -1$ -- wrap around the transverse $S^{p-1}$, giving rise to codimension $q$ defects in the dimensionally reduced description. These can similarly have topological endpoints on the defect $\calD$. Notice that, if instead $d-q < p -1$ the symmetry cannot act on $\calD$.\footnote{One way to interpret this is that all $\cL$ configurations ending on $\calD$ can be shrunk topologically.}

\subsection{SymTFT description of Defect Multiplets}
We now move to the SymTFT description of defect operators. As already explained in the Introduction, a defect $\calD$ of codimension $p$ belongs to a symmetry multiplet $\calL[S^{p-1}]$ described by a gapped boundary condition of the reduced SymTFT $\cZ(\cC)[S^{p-1}]$. Important information about the multiplet structure of $\calD$ is encoded in the topological interface $\calM_{\calL[S^{p-1}], \, \calL_{sym}[S^{p-1}]}$ between the reduced defect b.c. and the \soutn{topological} \rev{symmetry} one. This encodes the data of the higher Module category. We describe its salient features in \ref{sssec: calM}. We will \soutn{usually} \rev{often} denote this simply by $\calM$ as long as there is no risk for confusion.
After this, we move onto defect multiplets, which describe charged operators and domain walls on which the symmetry $\cC$ can act. This will be the content of \ref{sssec: vmult}. A complementary perspective, as well as some applications, will be given in \cite{CCboundaries}.

\vspace{2mm}\noindent \textbf{Boundary conditions}
\rev{Topological boundary conditions for $\cZ(\cC)$ are described by (higher) Lagrangian algebras $\calL$ of $\cZ(\cC)$ \cite{Kapustin:2010hk,Kapustin:2010if,Kong:2013aya}. Such objects are well characterized for Modular Tensor Categories, which correspond to a SymTFT description of a $1+1$d system via the Reshetikhin-Turaev construction. See \cite{Kaidi:2021gbs,Benini:2022hzx} for reviews aimed at physicists.}
Intuitively, a Lagrangian algebra $\calL$ is a maximal set of defects (and their junctions) which is mutually undetectable. Maximality implies that all other topological objects are detected (e.g. through braiding) by $\calL$.
Decorating the theory with a fine-enough mesh of $\calL$ describes a generalized gauging procedure leaving behind a trivial (invertible) theory. This is usually done by choosing a fine triangulation of spacetime $Y$. Mutual undetectability assures that the final answer does not depend on the choice of triangulation, by requiring invariance under the appropriate Pachner moves. 
Gauging the symmetry in half spacetime gives rise to a topological domain wall between the trivial theory and the starting one, which describes the gapped boundary condition. Given the gapped boundary condition, the structure of $\calL$ can be reconstructed by studying the ways in which bulk objects are allowed to terminate on it.

The definition of defect charges requires topological boundary conditions on \soutn{special} \rev{selected} manifolds with the topology $Y = Y_{d-p+2} \times S^{p-1}$, which we specify by a Lagrangian algebra $\calL[S^{p-1}]$ of the dimensionally-reduced SymTFT. This is a much larger set than that of gapped boundary conditions on generic manifolds, as the dimensional reduction \soutn{will trivialize various} trivializes several un-detectability constraints. A paradigmatic example is given by Chern-Simons theory. For concreteness consider $U(1)_k$, with $k$ not a perfect square. This theory does not admit gapped boundary conditions \cite{Kapustin:2010hk}. However we can also consider its $S^1$ reduction. \soutn{This can be performed at the level of the action and leads} \rev{This corresponds} to the \soutn{standard} $1+1$d BF theory for $\abA=\bZ_k$, which has $k$ indecomposable topological boundary conditions. These are the topological line operators in the original CS theory, we will consider related examples in detail later.

\rev{Notice that the initial topological boundary condition $\calL_{sym}$, which is valid on any manifold, will always descend to a reduced b.c. $\calL_{sym}[S^{p-1}]$. Its Lagrangian algebra object can be obtained from the higher dimensional one by compactification.}

\subsection{The junction $\calM_{\calL[S^{p-1}], \calL_{sym}[S^{p-1}]}$: symmetry breaking and $\cC$ action}\label{sssec: calM}
\begin{figure}
    \centering
    \begin{tikzpicture}[scale=0.75]
         \draw[color=red, fill=white!90!red, opacity=0.75] (-3,0) -- (-1,1) -- (-1,3) -- (-3,2) -- cycle;
  
     \draw[color=blue, fill=white!90!blue, opacity=0.75] (-1,1) -- (-1,3) node[above] {$\calM$} -- (1,3) -- (1,1) -- cycle;
     \draw[red, rotate around={30:(-2,1.5)},thick] (-2,1.5) ellipse (0.5 and 0.4);
     \draw  plot[smooth,tension=1.5] coordinates{(-2,1.5) (-0.5,1.25) (0,2) };
     \draw[fill=black] (-2,1.5) circle (0.05);  \draw[fill=black] (0,2) circle (0.05); 
      \node[below, red] at (-3,0) {$\calL_{sym}$};
     \node[above, blue] at (1,3) {$\calL[S^{p-1}]$};
     \node[above] at (-0.5,1.25) {$\lambda$};

\node[red] at (-2,2.15) {$\cL$};
\node at (2,2) {\Large$\overset{\text{lift}}{\leadsto}$};
\begin{scope}[shift={(6,0)}]
    \draw[color=red, fill=white!90!red, opacity=0.75] (-3,0) -- (-1,1) -- (-1,3) -- (-3,2) -- cycle;
  
     \draw[color=blue, fill=white!90!blue, opacity=0.75] (-1,1) -- (-1,3) node[above] {$\calM$} -- (1,3) -- (1,1) -- cycle;
     \draw  plot[smooth,tension=1.5] coordinates{(-2,1.5) (-0.5,1.25) (0,2) };
     \draw[fill=black] (-2,1.5) circle (0.05);  \draw[fill=black] (0,2) circle (0.05); 
     \node[below, red] at (-3,0) {$\calL_{sym}$};
     \node[above, blue] at (1,3) {$\calL[S^{p-1}]$};
     \node[above] at (-0.5,1.25) {$\lambda$};
     \draw (-0.8,1.25) arc (0:350:0.4 and 0.5);
     \node at (2.15,2) {\Large$\overset{\text{project}}{\leadsto}$};
     \begin{scope}[shift={(6.25,0)}]
    \draw[color=red, fill=white!90!red, opacity=0.75] (-3,0) -- (-1,1) -- (-1,3) -- (-3,2) -- cycle;
  \draw[color=blue, fill=white!90!blue, opacity=0.75] (-1,1) -- (-1,3) node[above] {$\calM$} -- (1,3) -- (1,1) -- cycle;
   \draw  plot[smooth,tension=1.5] coordinates{(-2,1.5) (-0.5,1.25) (0,2) };
     \draw[fill=black] (-2,1.5) circle (0.05);  \draw[fill=black] (0,2) circle (0.05); 
     \node[below, red] at (-3,0) {$\calL_{sym}$};
     \node[above, blue] at (1,3) {$\calL[S^{p-1}]$};
     \node[above] at (-0.5,1.25) {$\lambda$};
     \draw[blue,thick] (0,1.6) arc (-90:-440:0.4 and 0.4);
     \node[above] at (0,2.4) {$v$};
     \end{scope}
\end{scope}
     \end{tikzpicture}
    \caption{Sliding a symmetry operator across the order parameter $\lambda$ to prove that the symmetry is \rev{broken} \soutn{SSB} by the defect.}
    \label{fig: sliding}
\end{figure}
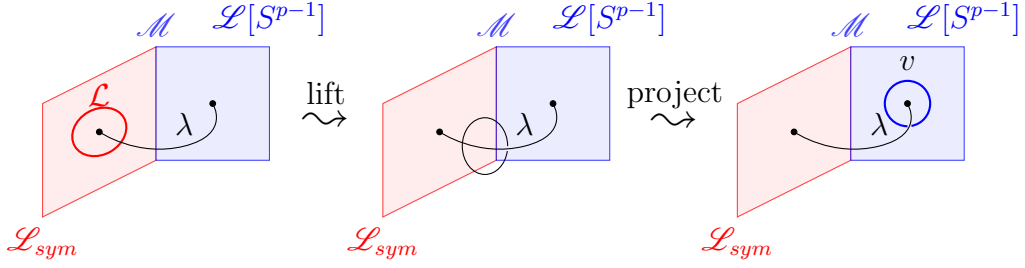

It is expected that, in the absence of bulk local topological operators, all boundary conditions $\calL[S^{p-1}]$ allow for \rev{topological} junctions $\calM$ with the $\calL_{sym}[S^{p-1}]$ boundary.\footnote{This is certainly true for Dijkgraaf-Witten type theories for which these junctions describe discrete gauging of subgroups of the $\cC$ symmetry, describing the Morita equivalence class of $\cC$.} A \rev{topological} junction $\calM_{\calL[S^{p-1}], \calL_{sym}[S^{p-1}]}$ \soutn{will describe} \rev{describes the} defect charge associated to $\calL[S^{p-1}]$. \soutn{this is akin to the concept of representation space for a group} \rev{This should be contrasted to case in which the symmetry is group-like, and $\calM$ essentially encodes the a representation space on which the group acts.}. 

The junction will in general not be indecomposable, that is, \soutn{on given topologies} \rev{upon compactification on given topologies} it will \soutn{allows for} \rev{host} local topological operators. \rev{Alternatively, the interface Hilbert space on certain spatial slices is degenerate.}

We explain below that this splitting signals \soutn{SSB} symmetry breaking \soutn{of some of the symmetry} by the defect. \rev{In the group theory example,} a simple component $\calM_a$ of $\calM$ is akin to a basis vector in the representation space. \rev{We presently describe the order parameters for such symmetry breaking and how to extract them from the SymTFT.}

Given the algebras $\calL[S^{p-1}]$ and $\calL_{sym}[S^{p-1}]$ define their intersection:
\be
\calL[S^{p-1}] \cap \calL_{sym}[S^{p-1}] = \left\{ \begin{array}{c}\text{Objects in $\cZ(\cC)[S^{p-1}]$ that can terminate} \\  \text{topologically on both $\calL$ and $\calL_{sym}$} \end{array} \right\} \, .
\ee
According to the \soutn{standard} SymTFT description, objects in the intersection are charged under the $\cC\rev{[S^{p-1}]}$ symmetry. \soutn{Suppose now that we} \rev{Let us} pick an object $\lambda$ of codimension $q$ in $\cZ(\cC)[S^{p-1}]$ in the intersection, \rev{and allow it to end of both topological boundary conditions simultaneously}. Performing the wedge compactification as in Figure \ref{fig: wedgecomp} \rev{and then considering the associated TFT $\cT_{\calL[S^{p-1}]}$ on a manifold $Y_{d-p+1} = S^{d-p-q+1}\times \Sigma_{q}$ leads to a degenerate Hilbert space $\dim\left( \cH_{ S^{d-p-q+1} \times \Sigma_q} \right) > 1$. Its states (also called universes) can be described by finding an idempotent basis $\pi_a$ generated from $V_\lambda \equiv  \lambda[S^{d-p-q+1}]$:}
\soutn{and then a further compactification on $S^{d-p-q +1}$ --so that the bulk operator becomes a line-- gives rise to a local topological operator $V_\lambda$ in the associated TFT.\footnote{More precisely, there can be multiplicities for there operators owning to the dimensionality of the junction space.} This implies that the Hilbert space $\cH_{\Sigma \times S^{d-p-q+1}}$ on $\Sigma \times S^{d-p-q+1}$ is not one-dimensional. Its states (also called universes) are obtained by rounding up all the operators $V_\lambda$ and performing a change of basis such that their algebra takes the form}
\be
\pi_a \pi_{b} = \delta_{ab} \, \pi_a \, . 
\ee
Projection on an \rev{eigenstate of $\pi_a$}  \soutn{state} describes an indecomposable component $\calM_a$ of the junction $\calM_{\calL[S^{p-1}], \calL_{sym}[S^{p-1}]}$. As these objects are charged under the \rev{dimensionally reduced} $\cC\rev{[S^{p-1}]}$ symmetry, they act as order parameter for the $\cC\rev{[S^{p-1}]}$ breaking, thus forbidding the defect $\calD$ from being \rev{strongly} symmetric. We thus conclude that
\begin{center}
\textit{
    Indecomposable components $\rev{\calM_a}$ of $\calM_{\calL[S^{p-1}], \rev{\calL_{sym}[S^{p-1}]}}$ on $\Sigma \times S^{d-p-q}$ are described by codimension $q$ bulk operators $\lambda$ ending on both $\calL[S^{p-1}]$ and $\calL_{sym}$, \rev{which act as order parameters for the broken bulk symmetry}.}
\end{center}

We now tie this observation with our previous definition of \soutn{spontaneously} broken defect symmetry. Consider a line $\lambda \in \calL[S^{p-1}] \cap \calL_{sym}$ and a $\cC[S^{p-1}]$ symmetry defect $\cL$ acting on it via linking on $\calL_{sym}[S^{p-1}]$ \rev{with a nontrivial charge}. At least one such defect exists due to $\lambda$ describing a bulk charged object \rev{under $\cC$}.

We can lift \soutn{this defect} \rev{the defect $\cL$} into the bulk, albeit in a non-unique manner, by considering the pre-image $\mu$ under the map $p_{\calL_{sym}\rev{[S^{p-1}]}}$ projecting a bulk operator to a boundary operator. We then slide the the topological operator onto the second gapped boundary $\calL[S^{p-1}]$, using the projection map $p_{\calL[S^{p-1}]}$. As the \soutn{symmetry} \rev{linking} action was nontrivial, it must be that $p_{\calL[S^{p-1}]}(\mu) \neq \unit$. Thus the symmetry operator $\cL$ cannot \soutn{terminate topologically on} \rev{be absorbed by} $\calM_{\calL[S^{p-1}], \calL_{sym}[S^{p-1}]}$ and it must describe a symmetry broken by the defect. The setup is shown in Figure \ref{fig: sliding}. This ties up our definition of broken symmetry with the standard SymTFT picture \cite{Bhardwaj:2023fca}.

The $\cC[S^{p-1}]$ symmetry acts on the defect $\calD$ through its topological endpoints $e_\cL$. These must satisfy various consistency conditions -- which we do not report here -- but coincide with those described on the defect worldvolume. These are packaged in the data of an higher module category over $\cC[S^{p-1}]$. The SymTFT does not give a lot of mileage in determining them, so we will not describe them in detail in this Note.

\subsection{Defect operator multiplets \rev{from SymTFT}} \label{sssec: vmult}
Next we describe (\rev{possibly twisted}) defect operator multiplets \soutn{and twisted defect operator multiplets}. Consider topological operators $v$ confined on the defect boundary $\calL[S^{p-1}]$. These form an (higher) category which we denote by $\cC^*[S^{p-1}]_{\calM}$. \rev{Physically, $\cC^*_{\calM}$ describes the ``dual" symmetry obtained form $\cC$ by generalized gauging. The specific gauging procedure is encoded in $\calM$ \cite{Choi:2023xjw,Diatlyk:2023fwf}.}
The interface $\calM$ implements a map between the category $\cC[S^{p-1}]$ and $\cC^*[S^{p-1}]_{\calM}$. \soutn{This map can be loosely interpreted as a generalized gauging transformation.}
\soutn{The generic configuration} \rev{A generic configuration of topological defects in this setting} is described by a defect doublet $(\cL, \, v)$ meeting at the interface $\calM$. We denote their junction by $e_{\cL, v}$. \soutn{We have already studied the case $v=1$ in the previous subsection to define the defect symmetry.} \rev{The special case $v=\unit$, which we have analyzed in the previous subsection, describes symmetries preserved by the defect.}\footnote{\rev{The existence of the junction will just ensure that the defect is \emph{weakly} symmetric.}}

\begin{figure}
    \centering
\begin{tikzpicture}[baseline={(0,-0.25)}][scale=0.75]
    \filldraw[color=white!90!red] (0,0) node[left,red] {$\calL_{sym}$} -- (0,2) -- (2,2) -- (2,0) --cycle;
       \filldraw[color=white!90!blue] (2,0) -- (2,2) -- (4,2) -- (4,0) node[right,blue] {$\calL[S^{p-1}]$} --cycle;
       \draw[black,thick] (2,0)--(2,2);
       \draw[red,thick] (0,1) -- (2,1); \draw[blue,thick] (2,1) -- (4,1);
       \node[red,above] at (1,1) {$\cL$}; \node[blue,above] at (3,1) {$v$};
\draw[fill=black] (2,1) circle (0.05);
\end{tikzpicture} \hspace{2cm}
    \begin{tikzpicture}[scale=0.75]

\node[right] at (3,2) {$=$};

     \draw[color=red, fill=white!90!red, opacity=0.75] (-2,0.5) -- (-1,1) -- (-1,3) -- (-2,2.5) -- cycle;
     \draw[red,thick] (-2,1.5) node[left] {$\cL$} -- (-1,2);
     \draw[color=blue, fill=white!90!blue, opacity=0.75] (-1,1) node[below] {$\calM_a$} -- (-1,3) node[above] {$\calM_b$} -- (2,3) -- (2,1) -- cycle;
     \draw[blue,thick] (-1,2) -- (2,2); \node[blue,above] at (0.5,2) {$v$}; \draw[fill=black] (-1,2) circle (0.05); \draw[fill=black] (2,2) circle (0.05);
      \draw[color=white!70!gray, fill=white!90!gray, opacity=0.75] (1,0.5) -- (2,1) -- (2,3) -- (1,2.5) -- cycle;
      \draw[blue, line width =2,opacity=0.75] (2,1) -- (2,3) node[above,gray] {$\partial X_{rel}$};
      \begin{scope}[shift={(3.5,0)}]
              \draw[color=white!70!gray, fill=white, opacity=0.75] (1,0.5) -- (2,1) -- (2,3) -- (1,2.5) -- cycle;   
                \draw[red,thick] (1,1.5) node[left] {$\cL$} -- (2,2);
\draw[gray, line width =2] (2,1) node[below] {$B_a$} -- (2,3) node[above] {$B_b$};
\draw[fill=black] (2,2) node[right] {$\phi^v_{a b}$} circle (0.05);
      \end{scope}
    \end{tikzpicture}
    \caption{Left, $\calM$ as a map between $\cC[S^{p-1}]$ and $\cC^*[S^{p-1}]_{\calM}$. Right, interval compactification of a generic map between $\cC$ and $\cC^*_{\calM}$.}
    \label{fig: defmulttwist}
\end{figure}
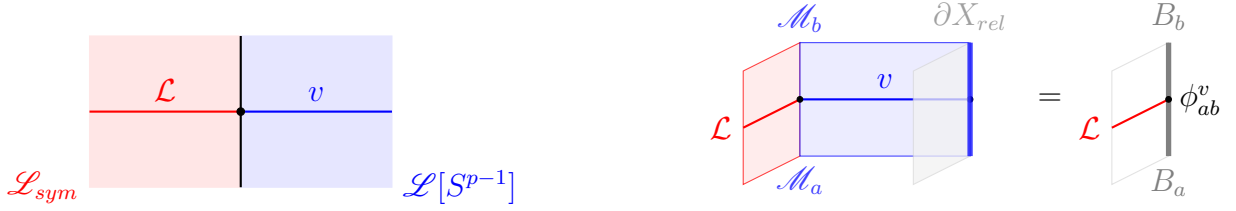

Performing interval compactification describes a topological defect $\cL$ ending on a non-topological operator $\phi^v$ \rev{localized on $\calD$}, the setup is shown in Figure \ref{fig: defmulttwist}. A special instance of this is when $\cL=1$ and the interval compactification describes a genuine defect \rev{operator} (multiplet). Such multiplet is charged under $\cC$, essentially replicating \eqref{eq: defectsymm}.

Lines in $\cC^*_\calM$ have a natural fusion structure, which describes the analogue of the tensor product for \soutn{standard} representations. This is extremely useful, as it gives a straightforward manner to prove selection rules for defect correlators. This will be used in \cite{Copetti:2024dcz} to implement S-matrix bootstrap for (1+1)d integrable systems with non-invertible symmetries.

All in all, taking into account both symmetries, $\calM$ is upgraded to be an element of an (higher) bi-module category. This mathematical object describes at the same time the action of the symmetry $\cC\rev{[S^{p-1}]}$ on the defect as well as the allowed defect multiplets.

\subsection{Defect OPE} Finally, let us describe constraints imposed by the symmetry $\cC$ on the defect OPE. 
Consider a bulk operator $\cO$, which can either be local or extended, which carries a charge $\lambda \, \in \, \cZ(\cC)$ under the symmetry. It is natural to consider its defect OPE:\footnote{For concreteness we write this formula with a local operator in mind. The coordinates $x$ describe the defect worldvolume placed at $z=0$.}
\be
\cO_{x,z} \overset{z \to 0}{\simeq} \sum_{\phi} z^{-\Delta_\cO + \Delta_\phi} \, b_{\cO \phi} \, \phi(x) \, .
\ee
We want to understand which defect operator multiplets $\phi^v$ are allowed to appear in such OPE. We start by considering the charge $\lambda$ stretching in the bulk in the presence of the defect $\calL[S^{p-1}]$ and perform the compactification. We then push the topological operator $\lambda[S^{p-1}]$ on the defect boundary and employ the projection map $p_{\calL[S^{p-1}]}$ to describe its boundary OPE, which takes the form:
\be
p_{\calL[S^{p-1}]} \left( \lambda[S^{p-1}] \right) = \sum_v  n^{\calL[S^{p-1}]}_{\lambda \, v} \, v \, ,
\ee
where $n^{\calL[S^{p-1}]}_{\lambda \, v}$ denotes the number of inequivalent indecomposable junctions between $\lambda[S^{p-1}]$ and $v$. Notice that, if a defect is not symmetric, a charged object $\lambda$ can be mapped into the neutral ($v=\unit$) defect operator multiplet.

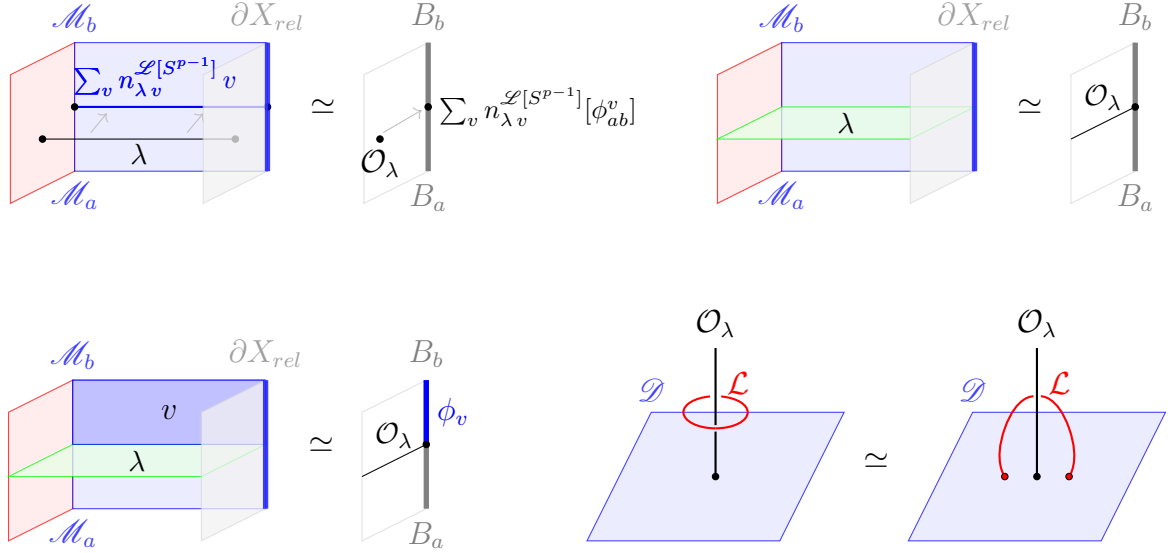
\begin{figure}
    \centering
\begin{tikzpicture}[scale=0.85]
    \node[right] at (2.5,2) {$\simeq$};

     \draw[color=red, fill=white!90!red, opacity=0.75] (-2,0.5) -- (-1,1) -- (-1,3) -- (-2,2.5) -- cycle;
    
     \draw[color=blue, fill=white!90!blue, opacity=0.75] (-1,1) node[below] {$\calM_a$} -- (-1,3) node[above] {$\calM_b$} -- (2,3) -- (2,1) -- cycle;
\draw[white!50!gray,->] (-0.75,1.6) -- (-0.5,1.9); \draw[white!50!gray,->] (0.75,1.6) -- (1,1.9);
     \draw[blue,thick] (-1,2) -- (2,2); \node[blue,above] at (0.25,2) {\footnotesize$ \sum_v  n^{\calL[S^{p-1}]}_{\lambda \, v} \, v$}; \draw[fill=black] (-1,2) circle (0.05); \draw[fill=black] (2,2) circle (0.05);
    \draw (-1.5,1.5) -- (1.5,1.5); \draw[fill=black] (-1.5,1.5) circle (0.05); \draw[fill=black] (1.5,1.5) circle (0.05); \node[below] at (0,1.6) {\small$\lambda$};
    
      \draw[color=white!70!gray, fill=white!90!gray, opacity=0.75] (1,0.5) -- (2,1) -- (2,3) -- (1,2.5) -- cycle;
\node[blue,above] at (0.25,2) {\footnotesize$ \sum_v  n^{\calL[S^{p-1}]}_{\lambda \, v} \, v$};
      
      \draw[blue, line width =2,opacity=0.75] (2,1) -- (2,3) node[above,gray] {$\partial X_{rel}$};
      
      \begin{scope}[shift={(2.5,0)}]
              \draw[color=white!70!gray, fill=white, opacity=0.75] (1,0.5) -- (2,1) -- (2,3) -- (1,2.5) -- cycle;   
\draw[gray, line width =2] (2,1) node[below] {$B_a$} -- (2,3) node[above] {$B_b$};
\draw[fill=black] (2,2) node[right] {\footnotesize$\sum_v  n^{\calL[S^{p-1}]}_{\lambda \, v} [\phi^v_{a b}]$} circle (0.05);
\draw[fill=black] (1.25,1.5) node[below] {$\cO_\lambda$} circle (0.05); 
\draw[white!50!gray,->] (1.3,1.6) -- (1.9,1.95);
      \end{scope}
\begin{scope}[shift={(11,0)}]
      \node[right] at (2.5,2) {$\simeq$};

     \draw[color=red, fill=white!90!red, opacity=0.75] (-2,0.5) -- (-1,1) -- (-1,3) -- (-2,2.5) -- cycle;
     \draw[color=blue, fill=white!90!blue, opacity=0.75] (-1,1) node[below] {$\calM_a$} -- (-1,3) node[above] {$\calM_b$} -- (2,3) -- (2,1) -- cycle;
     \draw[color=green, fill=white!90!green, opacity=0.75] (-2,1.5) -- (-1,2) -- (2,2) -- (1,1.5) --cycle; \node at (0,1.75) {\small$\lambda$};

       \draw[color=white!70!gray, fill=white!90!gray, opacity=0.75] (1,0.5) -- (2,1) -- (2,3) -- (1,2.5) -- cycle;
          \draw[blue, line width =2,opacity=0.75] (2,1) -- (2,3) node[above,gray] {$\partial X_{rel}$};
\begin{scope}[shift={(2.5,0)}]

          \draw[color=white!70!gray, fill=white, opacity=0.75] (1,0.5) -- (2,1) -- (2,3) -- (1,2.5) -- cycle;  
         
\node[above,black] at (1.5,1.78) {\small $\cO_\lambda$};
\draw[gray, line width =2] (2,1) node[below] {$B_a$} -- (2,3) node[above] {$B_b$};
 \draw (1,1.5) -- (2,2); \draw[fill=black] (2,2) circle (0.05);
\end{scope}
\end{scope}
      
\end{tikzpicture}
\\[1cm]
\begin{tikzpicture}[scale=0.85]
     \node[right] at (2.5,2) {$\simeq$};

     \draw[color=red, fill=white!90!red, opacity=0.75] (-2,0.5) -- (-1,1) -- (-1,3) -- (-2,2.5) -- cycle;
     \draw[color=blue, fill=white!90!blue, opacity=0.75] (-1,1) node[below] {$\calM_a$} -- (-1,3) node[above] {$\calM_b$} -- (2,3) -- (2,1) -- cycle;
     \draw[color=green, fill=white!90!green, opacity=0.75] (-2,1.5) -- (-1,2) -- (2,2) -- (1,1.5) --cycle; \node at (0,1.75) {\small$\lambda$};
      \draw[color=blue, fill=white!70!blue, opacity=0.75] (-1,2) -- (-1,3) -- (2,3) -- (2,2) --cycle;
\node at (0.5,2.5) {$v$};
       \draw[color=white!70!gray, fill=white!90!gray, opacity=0.75] (1,0.5) -- (2,1) -- (2,3) -- (1,2.5) -- cycle;
          \draw[blue, line width =2,opacity=0.75] (2,1) -- (2,3) node[above,gray] {$\partial X_{rel}$};

          \begin{scope}[shift={(2.5,0)}]
      \draw[color=white!70!gray, fill=white, opacity=0.75] (1,0.5) -- (2,1) -- (2,3) -- (1,2.5) -- cycle;  
         
\node[above,black] at (1.5,1.78) {\small $\cO_\lambda$};
\draw[gray, line width =2] (2,1) node[below] {$B_a$} -- (2,3) node[above] {$B_b$};
\draw[blue,line width=2] (2,2) -- (2,3); \node[blue, right] at (2,2.5) {$\phi_v$};
 \draw (1,1.5) -- (2,2); \draw[fill=black] (2,2) circle (0.05);
\end{scope}

\begin{scope}[shift={(7,0.5)}]
  \draw[color=blue,fill=white!90!blue,opacity=0.75] (0,0) -- (3,0) -- (4,2) -- (1,2) node[above] {$\calD$} -- cycle; 
\begin{scope}[shift={(2,1)}]
    \draw[thick] (0,0)  -- (0,0.7); \draw[thick] (0,0.8) -- (0,2) node[above] {$\cO_\lambda$};
  \draw[red,thick] (0, 0.75) arc(-90:-260: 0.5 and 0.25);  \draw[red,thick] (0, 0.75) arc(-90:80: 0.5 and 0.25);
   \node[red] at (0.35,1.45) {$\cL$};
\draw[fill=black] (0,0) circle (0.05);
\node at (2.5,0.25) {$\simeq$};
  \end{scope}
  \begin{scope}[shift={(5,0)}]
      \draw[color=blue,fill=white!90!blue,opacity=0.75] (0,0) -- (3,0) -- (4,2) -- (1,2) node[above] {$\calD$} -- cycle; 
\begin{scope}[shift={(2,1)}]
    \draw[thick] (0,0)  -- (0,2) node[above] {$\cO_\lambda$};
 \draw[red,thick] plot[smooth, tension=1.5] coordinates{(-0.5,0) (-0.5,0.75) (92.5:1.25)};
 \draw[red,thick] plot[smooth, tension=1.5] coordinates{(0.5,0) (0.5,0.75) (87.2:1.25)};
 \draw[fill=red] (-0.5,0) circle (0.05);  \draw[fill=red] (0.5,0) circle (0.05);
 \node[red] at (0.35,1.45) {$\cL$};
\draw[fill=black] (0,0) circle (0.05);
  \end{scope}
  \end{scope}
\end{scope}

\end{tikzpicture}
    \caption{Top-left: the bulk to defect OPE of a charge $\lambda$. Top-right, a defect $\cO_\lambda$ ending on $\calD$. Bottom-left, a defect $\cO_\lambda$ mapped into a nontrivial multiplet $v$. Bottom right: using a symmetric defect $\calD$ to un-link the symmetry action, implying that charged operators cannot terminate on symmetric defects.}
    \label{fig: defect OPE}
\end{figure}

For extended defects, the defect OPE also describes which bulk defect can end (non-topologically) on $\calD$. Consider the setup described in Figure \ref{fig: defect OPE}. In the SymTFT bulk we describe a genuine defect by the green surface stretching horizontally and ending on $\calL_{sym}[S^{p-1}]$. To prescribe the endpoint on $\calD$ we also need to specify a surface multiplet $v$. Only if $v = \unit$ is allowed can the defect terminate. Otherwise, it will continue into a defect operator multiplet $(v, \partial v)$. 
If $\calD$ is \emph{symmetric}, then no charged bulk operator $\cO_\lambda$ is allowed to terminate, as \soutn{using symmetry} we can use $\calD$ to unwind any configuration of bulk symmetry defects $\cL$ linking $\cO$.
We thus conclude:

\begin{center}
    \textit{(Extended) operators $\cO_\lambda$ charged under $\cC$ may terminate on $\calD$ only if the defect spontaneously breaks the symmetry.}
\end{center}

Let us give an example. Consider Maxwell theory in $4d$ with a boundary and the electric 1-form symmetry $U(1)^{(1)}$ \cite{DiPietro:2019hqe}. It is natural to consider Dirichlet and Neumann boundary conditions for $A$. Under the Dirichlet boundary condition the 1-form symmetry is broken by the boundary. Indeed the Wilson lines can terminate on it freely. The Neumann boundary condition is symmetric \rev{under $U(1)^{(1)}$} and Wilson line from the bulk simply become dynamical boundary Wilson lines.

\subsection{Remarks}

\vspace{2mm} \noindent \textbf{Local operators and (-1)-form symmetries} Sometimes the $S^{p-1}$ reduction of the SymTFT $\cZ(\cC)[S^{p-1}]$ contains local topological operators. In this case, should one wish to describe indecomposable defects, Dirichlet boundary conditions must be imposed on them. If this is not done, the defect $\calD$ will also host local topological operators, giving rise to \emph{decomposition} into universes \cite{Hellerman:2006zs,Lin:2022xod}:
\be
\calD = \bigoplus_i \calD_i \, ,
\ee
obtained by transforming the local topological operators into an idempotent basis. This will be a recurring theme in many examples.

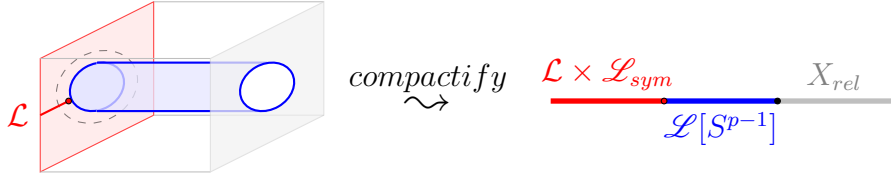
\begin{figure}
    \centering
    \begin{tikzpicture}[scale=0.75]
        \draw[color=red, fill=white!90!red, opacity=0.75] (-3,0) -- (-1,1) -- (-1,3) -- (-3,2) -- cycle;
      \draw[rotate around={30:(-2,1.5)},thick, blue, fill=white!90!blue] (-2,1.5) ellipse (0.5 and 0.4);
       \draw[rotate around={30:(-2,1.5)},dashed, opacity=0.5] (-2,1.5) ellipse (0.75 and 0.6);

\pgfmathsetmacro{\yA}{1.5 + sqrt((0.4*cos(30))^2 + (0.5*sin(30))^2 ) };
\pgfmathsetmacro{\yB}{1.5 - sqrt((0.4*cos(30))^2 + (0.5*sin(30))^2 ) };

\draw[color=white!90!blue, fill=white!90!blue, opacity=0.75] (-2,\yA) -- (1,\yA)-- (1,\yB)  -- (-2,\yB) -- cycle;
\draw[blue,thick] (-2,\yA) -- (1,\yA); \draw[blue,thick] (-2,\yB) -- (1,\yB);

   \draw[white!75!black] (0,0) -- (-3,0);    \draw[white!75!black] (2,1) -- (-1,1);   \draw[white!75!black] (2,3) -- (-1,3);   \draw[white!75!black] (0,2) -- (-3,2);     
   \draw[color=white!70!gray, fill=white!90!gray, opacity=0.75] (0,0) -- (2,1) -- (2,3) -- (0,2) -- cycle;   
         \draw[rotate around={30:(1,1.5)},thick, blue, fill=white] (1,1.5) ellipse (0.5 and 0.4);
         \draw[red, thick] (-3,1) node[left] {$\cL$} -- (-2.5,1.25); \draw[fill=red] (-2.5,1.25) circle (0.05);
         \node[right] at (2.25,1.5) {\Large$\overset{compactify}{\leadsto}$};
    \begin{scope}[shift={(6,0.25)}]
    
     \draw[red,line width=2] (0,1) -- (2,1); \draw[blue, line width=2] (2,1) -- (4,1) ; \draw[white!50!gray,line width=2] (4,1) -- (6,1);   
     \node[above,red] at (1,1) {$\cL \times \calL_{sym}$}; \node[below,blue] at (3,1) {$\calL[S^{p-1}]$}; \node[gray,above] at (5,1) {$X_{rel}$};
     \draw[fill=red] (2,1) circle (0.05); \draw[fill=black] (4,1) circle (0.05);
    \end{scope}     
    \end{tikzpicture}
    \caption{Left, description of a twisted sector defect in the SymTFT. Right, $S^{p-1}$ compactification (the dashed line on the left) gives a $(-1)$-form symmetry acting on the defect boundary condition.}
    \label{fig: twisted}
\end{figure}

Similarly, when local operators are present, the bulk also has codimension one magnetic operators, which generate a bulk zero-form symmetry. On $\calL_{sym}$, if dynamical, these describe a $(-1)$-form symmetry. In the language of defect operators its action on a boundary interface $\calM$ describes a twisted sector defect. See Figure \ref{fig: twisted} for a representation. 
For this reason, we will not treat invariance under $(-1)$-form symmetries on the same footing and will not be required in order to define symmetric defects.

\vspace{2mm} \noindent \textbf{The trivial defect} \rev{Let us describe the trivial defect $\unit_p$, which is present in any theory and at any codimension $p$.

The bulk symmetry $\cC[S^{p-1}]$ divides into two classes: symmetry operators which do not span any compactified direction, which we denote by $\cL_0$, and symmetry operators which have undergone compactification, denoted by $\cL_{comp}$. 
The former are broken by $\unit_p$, since parallel fusion obviously leads to the defect $\cL_0$, while the latter are preserved, as the (generalized) linking action on the identity defect must be trivial.  
Thus the interface $\calM$ must correspond to the regular module category (i.e. complete breaking) for $\cC[S^{p-1}]_0$ and a symmetry-preserving (in the strong sense) interface for $\cC[S^{p-1}]_{comp}$.
In the bulk SymTFT this has a simple description: the identity defect descends from the likewise-named identity defect $\unit_p$ of the bulk theory $\cZ(\cC)$. In the compactification process, all compactified topological defects $\lambda \in \cZ(\cC)[S^{p-1}]_{comp}$ must be allowed to terminate topologically on the $\calL_\unit[S^{p-1}]$ b.c., while uncompactified defects $\mu \in \cZ(\cC)[S^{p-1}]_{0}$ all have Neumann type boundary conditions on the interface. Notice that the two sets have nontrivial braiding relations between each other, but cannot braid within a single group.
We  consider the order parameters for this setup: as $\calL_{sym}[S^{p-1}]\cap\calL_\unit[S^{p-1}] = \calL_{sym}[S^{p-1}]_{comp}$. By the former remark, these are blind to the action of $\cC[S^{p-1}]_{comp}$ but braid nontrivially with the symmetry generators of $\cC[S^{p-1}]_0$, implying its breaking as anticipated.

The case $p=1$ of a codimension-1 defects deserves special attention. The $S^0$ compactification of $\cZ(\cC)$ is $\cZ(\cC\boxtimes \overline{\cC})$ by the folding trick. The symmetry boundary condition is described by the Lagrangian algebra $\calL_{sym} \otimes \overline{\calL_{sym}}$. A trivial interface preserves the diagonal symmetry $\cC_{diag} \subset \cC\boxtimes \overline{\cC}$ (although only in the weak sense!) and is implemented by the (canonical) diagonal Lagrangian algebra in $\cZ(\cC\boxtimes \overline{\cC})$: $\calL_{diag} = \bigoplus_{\lambda} (\lambda, \overline{\lambda})$. Indeed by construction only topological defects residing in projection on $\calL_{sym}[S^0]$ (namely the diagonal symmetry $\cC_{diag}$) are allowed to end topologically on the interface. Generic interfaces can similarly be described using the folding trick.
}

\soutn{By definition the trivial defect does not allow any symmetry defect to end upon it and thus --perhaps surprisingly-- spontaneously breaks the full categorical symmetry $\cC$. A boundary condition implementing the complete spontaneous breaking of $\cC$ is provided by the \emph{regular representation}, which corresponds to the universal choice $\calL_\calD[S^{p-1}] = \calL_{sym}[S^{p-1}]$. Decorating the interface between the two $\cL_{sym}$ boundaries by a symmetry defect describes the full set of topological defects of the bulk theory.
Thus, the set of topological defects forms a multiplet under the symmetry $\cC$. This is the (perhaps) familiar statement that $\cC$ is a module category over itself. }

\vspace{2mm} \noindent \textbf{A compact notation}
Given the remarks in \ref{sssec: calM} and \ref{sssec: vmult} we will associate to a defect charge $\calL[S^{p-1}]$ a diagram:
\bea
\begin{tikzpicture}
        \filldraw[color=white!90!red] (0,0) -- (3,0) -- (3,3) -- (0,3) -- cycle;  
      \filldraw[color=white!90!blue] (3,0) -- (6,0) -- (6,3) -- (3,3) -- cycle; 
      \draw[line width =2,gray] (3,0) -- (3,3);
      \node[red, below left] at (0,0) {$\calL_{sym}[S^{p-1}]$};
      \node[blue, below right] at (6,0) {$\calL[S^{p-1}]$};
      \draw[red] (0,1) -- (3,1); \node[above,red] at (1.5,1) {$\cL$}; \draw[fill=red] (3,1) circle (0.05); 
      \draw[red] (0,1.75) -- (3,1.75);  \node[above,red] at (1.5,1.75) {$\cS$}; \draw[blue] (3,1.75) -- (6,1.75); \node[blue,above] at (4.5,1.75) {$\phi_\cS$};
       \draw[blue] (3,2.5) -- (6,2.5); \node[blue,above] at (4.5,2.5) {$v$}; \draw[fill=blue] (3,2.5) circle (0.05); 
       \draw (2,0.5) arc (-180:0: 1 and 0.75); \draw[fill=black] (2,0.5) circle (0.05); \draw[fill=black] (4,0.5) circle (0.05); \node at (1.5,0.5) {$\lambda$}; \draw[fill=black] (3,1.75) circle (0.05); 
    
\end{tikzpicture}
\eea
Where -- from bottom to top -- we indicate:
\begin{enumerate}[i)]
    \item The symmetry-breaking parameters $\lambda$ describing the indecomposable components of $\calM$.
    \item The unbroken symmetry lines $\cL$. 
    \item The broken symmetry lines $\cS$, and their operator multiplets $\phi_\cS$.
    \item The local defect operator charges $v$.
\end{enumerate}

\rev{This will allow us to coincisely present the salient aspects of the following examples.}

\section{Examples}\label{sec: examples}
We now give some concrete applications of our formalism in various dimensions. 

\subsection{3d/2d correspondence}\label{ssec: 3d2d}
We start by exemplifying our methods by studying \soutn{the well known problem of} local charged operators in a $1+1d$ system. In this case the SymTFT is described by the Drinfeld center $\cZ(\cC)$ of the fusion category $\cC$. 
Boundary conditions on general manifolds are described by Lagrangian algebras $\calL$ \cite{Benini:2022hzx} in $\cZ(\cC)$ and \soutn{a genuine representation} \rev{the representation under which a genuine operator transforms} $\lambda = \ell$ by a line $\ell$ belonging to $\calL$. 

According to our discussion, these can also be described by boundary conditions for the reduced theory $\cZ(\cC)[S^1]$. The spectrum of operators in this theory is spanned by the lines $\lambda$ and by \rev{local} vertex operators:
\be
v_\lambda \equiv \lambda[S^1] \, ,
\ee
which encode the holonomies around the compactified $S^1$. An indecomposable Dirichlet boundary condition $\calL_\mu[S^1]$ is specified by consistent vevs for $v_\lambda$:
\be
\langle v_\lambda \rangle_\mu = B_{\lambda \, \mu} \, ,
\ee
satisfying the fusion algebra:
\be
B_{\lambda \, \mu} B_{\lambda' \, \mu} = \sum_{\lambda''} N_{\lambda \lambda'}^{\lambda''} \, B_{\lambda'' \mu} \, .
\ee
This is clearly just Verlinde's formula \cite{Verlinde:1988sn}, upon the identification:
\be
B_{\lambda \, \mu} = \frac{S_{\lambda \, \mu}}{S_{0 \, \mu}} \, .
\ee
Similarly one can check that:
\be
\lambda \times \calL_\mu[S^1] = \sum_{\mu'} N_{\lambda \mu}^{\mu'} \, \calL_{\mu'}[S^1] \, ,
\ee
by using the commutation relations between $v_\mu$ and $\lambda$. Thus we learn that Dirichlet boundary conditions correspond to simple lines in $\cZ(\cC)$.

Indecomposable boundary conditions in a 2d TFT are in correspondence with local idempotents $\pi_\mu$: $\pi_\mu \times \pi_\nu = \delta_{\mu\nu} \, \pi_\nu$ \cite{Huang:2021zvu} and in our case:
\be
\pi_\mu = \frac{1}{S_{0 \mu}} \sum_{\lambda} S_{\lambda \mu}^* v_\lambda  \, ,
\ee
give such a basis. We conclude that all other boundary conditions can be realized as linear combinations of the Dirichlet one. Now let us discuss when a Dirichlet boundary condition $\calL_\mu[S^1]$ can terminate on the $\calL_{sym}$ boundary.
To this end let us perform the circle reduction of the setup, for the symmetry boundary we have (see e.g. \cite{Kaidi:2021gbs,Benini:2022hzx}):
\be 
\calL_{sym}[S^1] = \bigoplus_\lambda  n_\lambda \, \calL_\lambda[S^1] \, ,
\ee
since the $\calL_\mu[S^1]$ boundary condition are indecomposable, an interface between the two exists only if
\be
n_\mu \neq 0 \, .
\ee
Thus recovering the usual SymTFT prescription.

\subsection{(Twisted) Dijkgraaf-Witten theory}
A second example the twisted DW theory for a $q$-form symmetry based on an abelian group $\abA$, with action:
\be
S = 2 \pi i \int_Y \biggl(  c_{d-q-1} \cup \delta b_{q+1} + \omega(b_{q+1}) \biggr)\, ,
\ee
where $c_{d-q-1} \in C^{d-q-1}(Y,\abA^\vee)$ and $b_{q+1} \in C^{q+1}(Y,\abA)$, $\cup$ is the cup product stemming from the canonical pairing $\abA \times \abA^\vee \to \bR/\bZ$ and $\omega$ represents a possible 't Hooft anomaly for $\abA$, so that, once $c_{d-q-1}$ is integrated out:
\be
\omega \ \in \ H^{d+1}(B^q \abA, U(1)) \, .
\ee
\rev{In this Subsection we will be mainly interested in examples where the twist trivialized after compactification. We study some selected examples where the twist remains nontrivial in the latter parts of this Section.}

\vspace{2mm} \noindent \textbf{Topological defects.} In the absence of twist, topological defects are described by the operators
\be
U_a = \exp\left( 2 \pi i a \int c_{d-q-1} \right) , \, \ \ \ V_\alpha = \exp\left( 2 \pi i \alpha \int b_{q+1} \right) \, .
\ee
If a twist is present, care is required in defining the magnetic defects $U_a$. In this case, \rev{the magnetic operators are usually non-genuine
\be \label{eq: magneticdef}
U_a[\Sigma] \, \exp\left( 2 \pi i (-)^{(q+2)(d-q-1)} a \int_{Y: \partial Y = \Sigma} \nu(b_{q+1}) \right) \, ,
\ee
where we interpret the integral as en element in $\abA^\vee$.
The required dressing is determined by $\omega$ to ensure gauge invariance. Explicitly consider $t(\lambda_q, b_{q+1})$ defined through}
\soutn{defining $t(\lambda_q, b_{q+1})$ through}
$\omega(b_{q+1} + \delta \lambda_q) - \omega(b_{q+1}) = \delta b_{q+1} \cup t(\lambda_q, b_{q+1})$ 
\soutn{and
requiring an invariant action gives the following transformation law for $c_{d-q-1}$:}
\rev{Gauge invariance requires the following transformation law for $c_{d-q-1}$:}
\be
\delta_\lambda c_{d-q-1} = - (-)^{(q+2)(d-q-1)} t(\lambda_q, \, b_{q+1}) \, .
\ee
\rev{Now introduce an inflow action $\nu$ for $t$ by}
\soutn{We can characterize $t(\lambda_q, b_{q+1})$ as the reduction of the original 't Hooft anomaly in a background with a nontrivial $b_{q+1}$ flux.\footnote{More precisely, given a flux $a \ \in \ \abA$ the reduced anomaly is $t(\lambda_q, \, b_{q+1})[a]$.} It can be absorbed by a new inflow action $\nu(b_{q+1})$ such that}
\be
\nu(b_{q+1} + \delta \lambda_q) - \nu(b_{q+1}) = \delta t(\lambda_q, b_{q+1}) \, .
\ee
The magnetic defect \rev{\eqref{eq: magneticdef}} thus is non-genuine \rev{but gauge invariant}.\footnote{In some limiting cases this mechanism allows to ``open" up the $V_\alpha$ defects, trivializing them. This has been studied in \cite{Hsin:2018vcg} and given a SymTFT perspective in \cite{Argurio:2024oym}.}
\soutn{\be
U_a \, \exp\left((-)^{(q+2)(d-q-1)} 2 \pi i a \int \nu(b_{q+1}) \right)
\ee
unless the anomaly cancels upon reduction.}
In some cases \cite{Kaidi:2022cpf,Antinucci:2022vyk,Kaidi:2023maf} an alternative description is possible, in which one retains local defects $\cU_a$ at the cost of making them non-invertible. 

\rev{To see this consider} a $d-q-1$-dimensional TFT $\cT_a$ with an anomalous $q$-form symmetry and anomaly $(-)^{(q+2)(d-q-1)}\nu(a)$, then the combination:
\be
\cU_a \equiv U_a \times \cT_a(b_{q+1}) \, ,
\ee
is a gauge-invariant, non-invertible defect. Notice that in both cases the electric defects are not mutually local -- either because non-genuine or because of the braiding with the TFT part -- and thus cannot be condensed \rev{at the same time}.

\rev{Consider for example the 2+1d twisted $\bZ_N$ theory, with twist $\omega(b) = \frac{1}{N^2} b \delta b$. We find $\nu(b) = \frac{1}{N^2} \delta b \equiv \frac{1}{N} \beta(b)$. This defect can be terminated by $\cT_a = \exp\left(i a/N^2 \int b \right)$, which recovers the well known description of magnetic defects in the twisted $\bZ_N$ theory \cite{Barkeshli:2014cna}. }

\vspace{2mm} \noindent \textbf{Boundary conditions}
The canonical Dirichlet boundary condition fixes $b_{q+1}$ at the boundary
and corresponds to the condensation of genuine magnetic defects $V_\alpha$, that is:
\be
\calL_{0} = \left\{ (0, \alpha) \, , \ \ \ \alpha \, \in \, \abA^\vee \right\} \, .
\ee
\rev{We will use this as a symmetry boundary condition $\calL_{sym} = \calL_0$.}
The electric defects $U_a$ -- once pulled-back on the boundary -- become again genuine and describe the symmetry generators of $\abA^{(q)}$:
\be
\cL_a = U_a \vert_{\calL_{sym}} \, .
\ee
A generic gapped boundary condition \cite{Antinucci:2023ezl}\footnote{See also \cite{Cordova:2024jlk} for a very recent study using a lattice formulation.} is described, at the top level, by a subgroup $\abB \subseteq \abA$ with trivial anomaly:
\be
\omega(b_{q+1})= \soutn{0} \rev{d \eta}  \  \, , \ \ \text{if} \ \ b_{q+1} \, \in \, \abB \, ,
\ee
whose defects $U_b$ are genuine and a \rev{trivially linking} complement
\be \label{eq: NB}
N(\abB): \beta \, \in \, \abA^\vee: \, \beta[b] = 1 \ \ \ \forall b \in \abB\, .
\ee
This assures that all the genuine defects in the bulk braid nontrivially with the condensed objects. On the gapped boundary electric operators $U_a$ are dynamical, but are identified modulo $\abB$. Thus decomposing $a = b c$, with $b  \in \abB$, we have:
\be
U_a \vert_{\calL_{\abB}} = \cL_c \, ,
\ee
the $\cL_c$ are in general non-invertible. On the other hand, since $\abA^\vee/N(\abB) \simeq \abB^\vee$ magnetic operators in the quotient are dynamical. That is, if $\alpha = \beta  \gamma$:
\be
V_\alpha \textbf{} = v_\gamma \, .
\ee
This prescription is incomplete as it admits a choice of symmetry fractionalization for junctions. For simplicity assume that $d-q-1 > q + 1$. Then, at a generic junction of $d-2q$ defects $U_{b_i}$, $i=1,...,d-2q$, we can choose a symmetry fractionalization class $\xi(b_1, ... , b_{d-2q}) \in H^{d-2q}(\abB, N(\abB))$.\footnote{See \cite{Brennan:2022tyl,Delmastro:2022pfo} for a physics-oriented review of symmetry-fractionalization and some interesting physical applications.} This class might be subject to further consistency conditions depending upon dimensionality.

We denote this boundary condition by:
\be
\calL_{\abB, \xi} \, .
\ee
A special case is $d = 2 q$, where the magnetic complement takes the form:
\be 
(b, \beta \psi(b) )  \, , 
\ee
with $\psi(b)$ a group homomorphism: $\abA \to \abA^\vee$ such that:
\be
\beta[b] = 1 \ \text{and} \ \psi(b)[b'] =1 \, , \ \ \ \forall \ b, b' \in \abB\, .
\ee
The quantity $\psi(b)[b'] \equiv \chi(b,b')$ is a bicharacter which satisfies
\be
\chi(b,b')= (-)^{q} \chi(b',b) \, .
\ee
and encodes a choice of discrete torsion. The $\abA$-symmetric boundary condition corresponds the electric algebra $\calL_{\abA}$, which can only exist if $\omega=0$, i.e. the symmetry is anomaly-free. The reduced theory on $S^{p-1}$ depends nontrivially upon $p$ and $q$. Let us give an overview of some relevant cases. We will treat separately selected examples in which the anomaly does not trivialize in \ref{sssec: 3donef} and  \ref{ssec: KOZ}.

\vspace{2mm} \noindent \textbf{p-1 $>$ q + 1 and p-1 $>$ d-q -1} In this case the reduced theory is trivial. There is only one boundary condition corresponding to the trivial representation of the $\abA^{(q)}$ symmetry. One example is $d=4$, $p=4$ and $q=1$, which describes the action of one-form symmetry on local operators, which must be trivial as the defect can always be deformed away from the operator.

\vspace{2mm} \noindent \textbf{p-1 $>$ q + 1 and p-1 $\leq$ d-q -1}
The reduced theory is still of DW type, but now with a trivial anomaly:
\be
S[S^{p-1}] = 2 \pi i \int_{Y_{d-p+2}} c_{d-q - p} \cup \delta b_{q+1} \, ,
\ee
Importantly, we are now free to choose any subgroup $\abB \subseteq \abA$ to define a gapped boundary condition, in stark contrast with the case of codimension-one boundaries. The electric boundary condition:
\be
\calL_{\abA}[S^{p-1}] \, ,
\ee
describes a defect preserving the full $\abA$ symmetry of the bulk theory $X$, even tough the symmetry is anomalous.
The special case $p=d-q$ instead describes the electric defects as boundary conditions, via their holonomy:
\be
\exp(2 \pi i \, c_0) = \exp(2 \pi i \alpha) \, , \ \ \alpha \, \in \, \ \abA^\vee \, .
\ee
As the reduced theory is still DW, we expect all boundary conditions to admit junctions with $\calL_{sym}$, except for the special case of $p=d-q+1$ where the same remarks as \ref{ssec: 3d2d} apply.
From a QFT perspective, the appearance of $c_{d-q-p}$ in the dimensional reduction describes the fact that the $\abA^{(q+1)}$ symmetry acts on the defect by braiding $d-p-1$ directions around it and fusing along the remaining $d-q-1$ along the defect $\calD$. 

We also have non-trivial defect operators. Recall that the canonical boundary condition $\calL_{sym}$ is $\calL_{0}$. Let us consider the boundary condition $\calL_{\abB, \xi}[S^{p-1}]$.
Surface operators $v_\gamma$ can terminate on the interface with the symmetry boundary thanks to the Dirichlet boundary condition, giving rise to operators charged under the boundary symmetry $\abA$, of codimension $p +q -1$.
On the other hand, electric surfaces $\cL_c$ can pass through the interface. Consequently, only operators $\cL_b$ that generate the $\abB$ subgroup of $\abA$ can terminate on the defect from the symmetry boundary. From the bulk braiding it follows that:
\be
\cL_{b}[v_\gamma] = \gamma(b) \, .
\ee
Thus the defect labelled by $\calL_{\abB, \xi}[S^{p-1}]$ is $\abB$-symmetric. This is also the group under which defect operators are charged.
We summarise this in Table \ref{tab: gwdefects} below. 
\begin{table}[h]
    \centering
    \begin{tabular}{|c||c|c|c|} \hline \rule[-1em]{0pt}{2.8em} 
       & Algebra  & Preserved symmetry  & Defect Charges \cr \hline \hline \rule[-1em]{0pt}{2.8em} 
   Boundary condition   & $\calL_{\abB, \bN(\abB)}[S^{p-1}]$  & $\abB \subset \abA$ & $\abB^\vee \simeq \abA^\vee/\bN(\abB)$ \cr \hline \rule[-1em]{0pt}{2.8em} 
    Objects & $\left( U_b, \, V_\beta \right)$ & $\cL_b = U_b \vert_{\calL_{\abB, \bN(\abB)}}$ & $v_\gamma = V_{\beta \gamma} \vert_{\calL_{\abB, \bN(\abB)}}$ \cr \hline
    \end{tabular}
    
\vspace{0.5cm}

\begin{tikzpicture}
      \filldraw[color=white!90!red] (0,0) -- (3,0) -- (3,3) -- (0,3) -- cycle;  
      \filldraw[color=white!90!blue] (3,0) -- (6,0) -- (6,3) -- (3,3) -- cycle; 
      \draw[line width =2,blue] (3,0) -- (3,3);
      \node[red, below left] at (0,0) {$\calL_{sym}$};
      \node[blue, below right] at (6,0) {$\calL_{\abB, \xi}[S^{p-1}]$};
      \draw[red] (0,1) -- (3,1); \node[above,red] at (1.5,1) {$\cL_b$}; \draw[fill=red] (3,1) circle (0.05); 
      \draw[red] (0,1.75) -- (3,1.75);  \node[above,red] at (1.5,1.75) {$\cL_a$}; \draw[blue] (3,1.75) -- (6,1.75); \node[blue,above] at (4.5,1.75) {$\phi_a$};
       \draw[blue] (3,2.5) -- (6,2.5); \node[blue,above] at (4.5,2.5) {$v_\gamma$}; \draw[fill=blue] (3,2.5) circle (0.05); 
       \draw (2,0.5) arc (-180:0: 1 and 0.75); \draw[fill=black] (2,0.5) circle (0.05); \draw[fill=black] (4,0.5) circle (0.05); \node at (1.5,0.5) {$V_\beta$}; \draw[fill=black] (3,1.75) circle (0.05);
    \end{tikzpicture}
    \caption{Structure of the $\calL_{\abB, \, \xi}$ defect multiplet in the DW theory, the Figure follows the notation of Section \ref{sec: reps}.}
    \label{tab: gwdefects}
\end{table}

\subsection{$2+1$d, Anomalous 1-form symmetry}\label{sssec: 3donef} An example where the nontrivial 't Hooft anomaly matters is the SymTFT for a one-form symmetry in a 3d theory, we consider for concreteness $\abA^{(1)}=\bZ_N$:
\be
S = 2 \pi i \int  c \cup \delta b + \frac{p}{2} \fP(b) \, .
\ee
With $\fP$ the appropriate generalization of the Pontryagin square operation to open chains \cite{Benini:2018reh}. If $\gcd(p,N)=1$ the SymTFT is invertible, and there is only a Dirichlet boundary condition for $b$. 
The set of operators are \cite{Hsin:2018vcg}:
\be
V_r = \exp\left(2 \pi i r \int_\gamma b \right) , \, \ \  U_n = \exp\left(2 \pi i n \int_\gamma c  + 2 \pi i p n \int_{\Sigma : \, \partial \Sigma = \gamma} b\right) \, .
\ee
Line operators $U_n=\exp(2 \pi i n \int c)$ while not genuine implement the one-form symmetry on $\calL_{sym}$:
\be
U_n \vert_{\calL_{sym}} = \cL_n \, .
\ee
\vspace{2mm} \noindent \textbf{Line defects} A 3d theory typically hosts nontrivial line defects $\calD$, which we describe by the dimensional reduction:
\be
S[S^{1}] = 2 \pi i \int \phi \cup \delta b + c \cup \delta a + p \, a \cup b \, , \ \ \ \phi = \int_{S^1} c \, , \ a= \int_{S^1} b \, .
\ee
A simple line operator requires a Dirichlet boundary condition for $\phi$:
\be \exp(2 \pi i \phi) = \exp(2 \pi i q/N) \ee
which specifies its one-form symmetry charge. If $\gcd(N,p)=1$, the mixed anomaly term forces us to choose Dirichlet boundary conditions also for $a$. Thus on a simple line defect $\calD$ with anomalous one-form symmetry $c$ and $b$ are dynamical and describe a domain wall $\cL_n$.
We denote this boundary condition by $\calL_q[S^1]$.
Since $c$ is dynamical the one-form symmetry is spontaneously broken by the line, reflecting the fact that:
\be
\cL_n \times \calD \neq \calD \, ,
\ee
as the fusion product must carry one-form symmetry charge $q+ np \neq q \mod n$. 
If $\gcd(p,N) = k \neq 1$ and $N=kr$ we can impose Neumann boundary conditions on the lines $V_{r s} = \exp(2\pi i r s \int a) $, $s=0, ... k-1$, or, equivalently, Dirichlet for $U_{r s}$. 
The $\bZ_k$ one-form symmetry is then unbroken on the defect the lines:
\be
v_s = V_{r s} \vert_{\calL_{q}[S^1]} \, ,
\ee
describe local operators living on the line $\calD$ which are charged under the preserved $\bZ_k$ one-form symmetry.
We summarize the results in Table \ref{tab: anom1f}. 
\begin{table}[h]
    \centering
    \begin{tabular}{|c||c|c|c|} \hline \rule[-1em]{0pt}{2.8em} 
  Case &   Algebra (Objects) & Preserved $\abA^{(1)}$ & Charged operators  \cr \hline \hline \rule[-1em]{0pt}{2.8em} 
   $\gcd(N,p)=1$  & $\left(e^{2\pi i \phi}, e^{2 \pi i \int a} \right)$  & $\varnothing$  & $\varnothing$ \cr \hline \rule[-1em]{0pt}{2.8em} 
   $\gcd(N,p)=k$  & $\left(e^{2\pi i \phi}, e^{2 \pi i k \int a} , e^{2 \pi i r \int c} \right)$ & $\bZ_k$ & $v_s = V_{r s} \vert_{\calL_{q}[S^1]}$ \cr \hline
    \end{tabular}

\vspace{0.5cm}

\begin{tikzpicture}
     \filldraw[color=white!90!red] (0,0) -- (3,0) -- (3,3) -- (0,3) -- cycle;  
      \filldraw[color=white!90!blue] (3,0) -- (6,0) -- (6,3) -- (3,3) -- cycle;
      \filldraw[color=white!70!blue] (3,1.5) -- (6,1.5) -- (6,3) -- (3,3) -- cycle;
      \draw[line width =2,blue] (3,0) -- (3,3);
      \node[red, below left] at (0,0) {$\calL_{sym}$};
      \node[blue, below right] at (6,0) {$\calL_{q}[S^1]$};
        \node[blue, above right] at (6,3) {$\calL_{q + n}[S^1]$};
     \draw[fill=black] (2,0.5) circle (0.05); \draw[fill=black] (4,0.5) circle (0.05); \node at (1.0,0.5) {$\phi = \text{free}$}; \node at (4.75,0.5) {$\phi = q$};
     \draw[red] (0,1.5) -- (3,1.5); \draw[blue] (3,1.5) -- (6,1.5); \node[red,above] at (1.5,1.5) {$\cL_n$}; \node[above,blue] at (4.5,1.5) {$(\cL_n V_{pn})$};
     \draw[fill=black] (4,2.75) circle (0.05); \node at (5.2,2.75) {$\phi=q+pn$};
     \node[right] at (8,1.5) {$\gcd(p,N)=1$};
\end{tikzpicture}

\vspace{0.5cm}

\begin{tikzpicture}
     \filldraw[color=white!90!red] (0,0) -- (3,0) -- (3,3) -- (0,3) -- cycle;  
      \filldraw[color=white!90!blue] (3,0) -- (6,0) -- (6,3) -- (3,3) -- cycle;
      \draw[line width =2,blue] (3,0) -- (3,3);
      \node[red, below left] at (0,0) {$\calL_{sym}$};
      \node[blue, below right] at (6,0) {$\calL_{q}[S^1]$};
     \draw[fill=black] (2,0.5) circle (0.05); \draw[fill=black] (4,0.5) circle (0.05); \node at (1.0,0.5) {$\phi = \text{free}$}; \node at (4.75,0.5) {$\phi = q$};
     \draw[red] (0,1.5) -- (3,1.5); \draw[fill=red] (3,1.5) circle (0.05);  \node[red,above] at (1.5,1.5) {$\cL_{rs}$};
     \draw[blue] (3,2.5) -- (6,2.5); \draw[blue] (3,2.5) circle (0.05);  \node[blue,above] at (4.5,2.5) {$v_{s}$};
     \node[right] at (7.85,1.5) {$\begin{array}{c} \gcd(p,N)=k  \\  N=kr \end{array}$};
\end{tikzpicture}

     \caption{Structure of charged defect multiplet for the anomalous 1-form symmetry.}
    \label{tab: anom1f}
\end{table}

\subsection{3$+$1d, KOZ Defects}\label{ssec: KOZ}
KOZ defects \cite{Kaidi:2021xfk} are non-invertible symmetries in $3+1$d. They are defined for -- say -- a $\bZ_N$ group by starting with a system having \soutn{$\bZ_n^{(0,1)}$ symmetries} \rev{$\bZ_N^{(0)} \times \bZ_N^{(1)}$ symmetry}  with mixed anomaly:
\be
I = \pi i p \int A \cup \fP(B) \, ,
\ee
and gauging the one-form symmetry\soutn{background $B$}. The zero-form symmetry defect is upgraded to a (non-invertible) KOZ defect $\cN$ satisfying:
\be
\cN \, U_n = U_n \, \cN = \cN \, , \ \ \ \cN \times \cN^\dagger = \text{Cond}(\bZ_N) \, ,
\ee
where $U_n$ are the (dual) one-form symmetry defects and $\text{Cond}(\bZ_N)$ is a condensation defect for the $\bZ_N$ symmetry \cite{Roumpedakis:2022aik}.\footnote{Explicitly, given a 3-surface $\Sigma$ and a one-form symmetry $\abA$:
\be \text{Cond}(\abA) = \sum_{\gamma \ \in \ H_2(\Sigma, \, \abA)} U(\gamma) \, . \ee
} 
The general fusion rules for these defects have been worked out in \cite{Copetti:2023mcq}.

\vspace{2mm} \noindent \textbf{SymTFT for KOZ defects}
KOZ type defects are described by the SymTFT \cite{Kaidi:2023maf}:
\be
S = 2 \pi i \int v_3 \cup \delta a_1 + c_2 \cup \delta b_2 + \frac{p}{2} a_1 \cup \fP(b_2) \, ,
\ee
where $a_1,v_3,c_2,b_2 \in C^{1,3,2,2}(Y, \bZ_N)$, respectively.\footnote{It is also possible to add a cubic anomaly term $2 \pi i \epsilon \int a_1 \cup \beta(a_1)^2$, this will not affect the discussion of defects below as its sphere reduction is trivial. We will also consider only the case $\gcd(p,n)=1$. Generalization is straighforward but tedious.}
The gauge transformations of the fields are \cite{Kaidi:2023maf}:
\bea
a_1 &\to a_1 + \delta \alpha_0  \, ,  \\
b_2 &\to b_2 + \delta \beta_1  \, ,   \\
c_2 &\to c_2 + \delta \gamma_1 + \alpha_0 \cup b_2 + \beta_1 \cup a_1 + \alpha_0 \cup \delta \beta_1 \, , \\
v_3 &\to  v_3 + \delta \nu_3 - p \beta_1 \cup b_2 - p \beta_1 \cup \delta \beta_1   \, .
\eea
Genuine topological defects are:
\be
W_n = \exp\left(2 \pi i \int a_1 \right) \, , \ \ \ U_\ell = \exp\left(2 \pi i \ell \int b_2 \right) \, , 
\ee
on the other hand defects for $v_3$ and $c_2$ are non-genuine:
\bea
&V_r = \exp\left(2 \pi i r \int_\gamma v_3 + \pi i r p \int_{\Sigma: \ \partial \Sigma = \gamma} \fP(b_2) \right) \, , \\ 
& U_s = \exp\left(2 \pi i s \int_\gamma c_2 - 2 \pi i p s \int_{\Sigma: \ \partial \Sigma = \gamma} a_1 \cup b_2 \right) \, .
\eea
both of them can be made genuine at the expense of introducing additional degrees of freedom on their worldvolume \cite{Antinucci:2022vyk,Kaidi:2023maf}. Their genuine avatars read:
\be
\cV_r = \exp\left( 2\pi i r \int v_3 \right) \, \cA^{N,r p}(b_2) \, , \ \ \ \ \ \  \cU_s = \exp\left(2 \pi i \int c_2 \right) \ \mathbf{Z}_N(a_1, b_2) \, ,
\ee
where $\cA^{N,p}$ is the minimal $\bZ_n$ theory of \cite{Hsin:2018vcg} and $\mathbf{Z}_N$ is the 2d $\bZ_N$ DW theory.

The KOZ symmetry is described by Dirichlet boundary conditions for $a_1$ and $c_2$. 
Defects $U_\ell$ implement the boundary 1-form symmetry while the bulk $\cV$ defect 
becomes the non-invertible KOZ defect:
\be
\cN = \cV_r \vert_{\calL_{sym}} \, .
\ee
The symmetry is anomaly-free, and the Fiber Functor is described by choosing Dirichlet boundary conditions for $v_3$ and $b_2$ instead.

\vspace{2mm} \noindent \textbf{Line defects (p $=$ 3)} We start by describing line defects with this symmetry. They correspond to an $S^2$ reduction:
\be
S[S^2] = 2 \pi i \int v_1 \cup \delta a_1 + b_0 \cup \delta c_2 + c_0 \cup \delta  b_2 + 2 \pi i p \int a_1 \cup b_0 \cup b_2 \, .
\ee
The new compactified non-genuine defects are
\bea
& V_r = \exp\left(2 \pi i r \int_\gamma v_1 - 2 \pi i r p \int_{\Sigma: \ \partial \Sigma = \gamma} b_0 b_2 \right) \, , \\ 
& U_s = \exp\left(2 \pi i s c_0 - 2 \pi i p s \int_{\gamma: \ \partial \Sigma = \text{p}} b_0 a_1 \right) \, .
\eea
Simple lines are dyons described by $(b_0, c_0) = (m, n)$ Dirichlet boundary conditions, which we denote by $\calL_{(m,n)}[S^2]$. The reduction of $\calL_{sym}$ is $\bigoplus_{m} \calL_{(m,0)[S^2]}$ so the electric boundary conditions $(m,0)$ describe genuine line operators.

Since $b_2$ is dynamical we further need to specify Dirichlet boundary conditions for $a_1$ unless $m=0$. The KOZ symmetry is thus SSB by these generic lines, and since the $V$ line is the boundary of an $U_{p m}$ surface, it acts as a domain wall between $(m,n)$ and $(m, n + p m)$ dyons. The simple case $n=0, p=1, m=1$ reproduces the mapping of the fundamental Wilson line $W$ into an 't Hooft line $T$ as shown by \cite{Kaidi:2021gbs}.\footnote{For $N=2$, KOZ is the same as the duality symmetry of \ref{sssec: su2}. However, in $SU(2)$ language, $(m,n)$ are not electric and magnetic charges, but rather electric and dyonic charges. The invariant dyon is $(0,1)$ and $(1,1)$ is an 't Hooft line. For $N>2$ this gives the right transformation law discussed in \cite{Choi:2022jqy,Cordova:2022ieu}.
}
Furthermore, since $b_2$ is dynamical, we find that this is a twisted sector line, as expected. 
On the other hand, if $m=0$, the line is free to either preserve or SSB the $\bZ_n$ KOZ symmetry in a standard manner.
\begin{table}
    \centering
   \begin{tikzpicture}
       \filldraw[color=white!90!red] (0,0) -- (3,0) -- (3,3) -- (0,3) -- cycle;  
      \filldraw[color=white!90!blue] (3,0) -- (6,0) -- (6,3) -- (3,3) -- cycle;
       \filldraw[color=white!70!red] (0,1.5) -- (3,1.5) -- (3,3) -- (0,3) -- cycle;
      \filldraw[color=white!70!blue] (3,1.5) -- (6,1.5) -- (6,3) -- (3,3) -- cycle;
      \draw[line width =2,blue] (3,0) -- (3,3);
      \node[red, below left] at (0,0) {$\calL_{sym}$};
      \node[red, above left] at (0,3) {$\calL_{sym}^{twisted}$};
      \node[blue, below right] at (6,0) {$\calL_{(m,n)}[S^2]$};
      \node[blue, above right] at (6,3) {$\calL_{(m,n +pm)}[S^2]$};
      \draw[fill=black] (2,0.5) circle (0.05); \draw[fill=black] (4,0.5) circle (0.05); \node[left] at (2,0.5) {$(b_0,c_0) = (\text{free},0)$}; \node[right] at (4,0.5) {$(b_0,c_0)= (m,n)$};
      \draw[red] (0,1.5) -- (3,1.5); \draw[blue] (3,1.5) -- (6,1.5);
\node[above] at (1.5,1.5) {$V$};
      \draw[fill=black] (4,2.5) circle (0.05); \node[right] at (4,2.5) {$(b_0,c_0)= (m,n + pm)$};
   \end{tikzpicture}
    \caption{Generic line multiplet under KOZ symmetry. The twisted symmetry b.c. is defined by the fusion product $\cL_p \times \calL_{sym}$.}
    \label{tab: KOZ multi}
\end{table}

Surface defects can also be analyzed in a similar manner, however since the analysis is quite cumbersome we do not attempt it here. Similar surface defects for duality symmetry will be discussed in detail in \ref{sec: examplesduality}.

\section{Defect multiplets for $(3+1)$d Duality Symmetry}\label{sec: examplesduality}
As the main application, we will consider defect multiplets under the $3+1$d self-duality symmetry \cite{Choi:2021kmx,Kaidi:2021xfk} for $p=2,3$ (surface and line defects, respectively). We will make extensive use of the perspective outlined in \cite{Antinucci:2023ezl,Cordova:2023bja} through the SymTFT description given in \cite{Kaidi:2022cpf,Antinucci:2022vyk}, which we both briefly review.\footnote{The methods used here can be extended in a straightforward manner to triality \cite{Thorngren:2021yso,Choi:2022zal} and G-ality \cite{Ando:2024hun,Lu:2024lzf} symmetries.}
\rev{These symmetries, especially in higher dimensions, naturally appear in a variety of critical systems, which can be either free (Maxwell theory) or strongly interacting ($\cN=4$ SYM at $\tau=i$. The methods outlined in this Section promise to be relevant for e.g. the classification of allowed surface (Gukov-Witten) operators and their transformation properties under the bulk symmetry.}

There will be two types of duality symmetry at play. The first, associated with invariance under the gauging of an abelian zero-form symmetry $\abA$ in $1+1$ or one-form symmetry $\abA^{(1)}$ in $3+1$d has been discussed at length in the literature \cite{Frohlich:2004ef,Frohlich:2006ch,Chang:2018iay,Choi:2021kmx,Kaidi:2021xfk}. In $3+1$d \rev{(which is the case we focus on in this presentation)} the symmetry is given by an $\abA^{(1)}$ invertible symmetry with generators $U_a$ and a duality defect $\cN$ satisfying:
\be
U_a \, \cN = \cN \, U_a = \cN \, , \ \ \ \cN \, \cN^\dagger = \text{Cond}(\abA) \, ,
\ee
where $\text{Cond}(\abA)$ is a condensation defect for the $\abA$ symmetry \cite{Roumpedakis:2022aik}. Notably \cite{Kaidi:2022uux}, this symmetry is relevant for $\cN=4$ SYM at $\tau=i$ (for simply laced gauge group), with $\abA$ the one-form symmetry group.

Its structure is determined by:
\begin{itemize}
\item An Abelian group $\abA$.
    \item A symmetric bicharacter $\chi$ on $\abA \times \abA$. 
    \item A discrete anomaly $\epsilon \in H^{d+1}(BG, U(1))$ where $G=\bZ_2, \, \bZ_4$ is the duality group.\footnote{Depending on the situation, it is more precise to think of this as an element of a bordism group.}.
\end{itemize}
These data were discussed in \cite{Thorngren:2019iar} in $1+1$d and \cite{Antinucci:2023ezl,Cordova:2023bja} in $3+1$d. The bicharacter provides an isomorphism between the symmetries $\abA$ and $\abA^\vee$ after the gauging while $\epsilon$ is a pure anomaly for the duality symmetry. In the present work we will consider the case of a trivial $\epsilon$.
The second duality symmetry is a three dimensional one, associated to the gauging of $\abA^{(0)} \times \abA^{(1)}$ \cite{Decoppet:2023bay,Cui:2024cav}. And is described by:
\begin{itemize}
    \item Two symmetric bicharacters $\chi_1, \chi_2$ on $\abA \times \abA$ providing isomorphisms between $\abA^{(0,1)}$ and ${\abA^{(1,0)}}^\vee$.
\end{itemize}
This will be the symmetry describing surface defects.
We now describe their symmetry TFT, with a focus on the $3+1$ dimensional case.

\subsection{SymTFT for Duality symmetry} To construct the SymTFT for the $3+1$d duality symmetry, we start with $\abA$ DW theory in 5d:
\be
S = 2 \pi i \int b_2 \cup \delta c_2 \, , \ \ b \ \in \ C^2(Y, \abA^\vee)\, , \  c \ \in \ C^2(Y, \abA) \, .
\ee
Denoting the dyonic operator $\exp\left(2 \pi i a \int b_2 + 2\pi i \alpha \int c_2 \right)$ by $(a, \alpha)$, a duality is generated by an isomorphism $\rho: \abA \to \abA^\vee$ via the transformation:
\be
S_\phi: \ (a, \, \alpha) \ \longrightarrow \ (-\phi^{-1}(\alpha) , \, \phi(a) ) \, .
\ee
This is equivalent to a symmetric non-degenerate bicharacter $\chi$ on $\abA \times \abA$ defined by:
\be
\chi(a,b) = \phi(a)[b] \, .
\ee
$S_\phi$ is a zero-form symmetry of the DW theory.
The SymTFT for the duality symmetry is obtained by gauging $S_\phi$ with discrete torsion $\epsilon$.
This will introduce a futher discrete gauge field, $\mathfrak{a} \in H^1(Y, G)$ and a corresponding magnetic (Gukov-Witten) defect $\mathfrak{N}$.

The canonical Dirichlet boundary condition $\calL_{sym}$ is \rev{a} Dirichlet \rev{b.c.} for $c_2$ and $\mathfrak{a}$. The boundary \soutn{pull-back} \rev{projection} of $\mathfrak{N}$ describes the duality defect:
\be
\cN = \mathfrak{N} \vert_{\calL_{sym}} \, .
\ee

\vspace{2mm} \noindent \textbf{Gapped boundary conditions} Gapped boundary conditions in $\cZ(\cC)$ can be deduced from those of the DW theory, together with the $S_\phi$ action on them. It is shown in \cite{Antinucci:2023ezl,Cordova:2023bja,Sun:2023xxv} that duality-invariant gapped boundary conditions in $\text{DW}(\abA)$ give rise to $\abA^{(1)}$ TFTs which are invariant under gauging $\abA^{(1)}$ with coupling $\chi$:
\be
S_\phi \cdot Z[B] = \# \sum_{b \ \in \ H^2(Y,\abA)} Z[b] \ \chi(b,B) \, .
\ee
When this happens the duality symmetry is \emph{Group Theoretical}, i.e. it can be recast as a 2Group after the appropriate discrete gauging.
If furthermore the invariant TFT is an SPT, then the duality symmetry is Anomaly-Free.

Gapped boundary conditions in DW theory are characterized by algebras $\calL_{\abB, \, \psi}$:
\be
\calL_{\abB, \, \psi} = \left\{ (b, \, \psi(b) \beta) , \, b \in \abB, \, \beta \in N(\abB) \right\}
\ee
with 
\be
\gamma(b,b') \equiv \psi(b)[b'] = \gamma(b',b) \, ,
\ee
a symmetric bicharacter and $N(\abB)$ defined in \eqref{eq: NB}. Denoting by $\text{Rad}_\psi$ the kernel of $\psi$, then according to \cite{Antinucci:2023ezl} $\calL_{\abB, \, \psi}$ is duality-invariant iff
\begin{enumerate}[i)]
\item $N(\abB) \simeq \text{Rad}_\psi$ and
\item The automorphism $\sigma = \phi^{-1} \psi: \abB/\text{Rad}_\psi \to \abB/\text{Rad}_\psi$ satisfies:
\be \label{eq: dualityinvfull}
\sigma^2 = -1 \ \ \ \text{and} \ \ \ \gamma(b,b') = \chi(\sigma(b), b') \, .
\ee
\end{enumerate}
Furthermore, the duality symmetry is anomaly-free iff
\be
\calL_{\abB, \, \psi} \cap \calL_{sym} \equiv \calL_{0} = \{1\} \, ,
\ee
that is $N(\abB) = 0$ and $\abB =\abA$.

When the duality-invariant algebra exists, one is free to impose Dirichlet boundary conditions on $\mathfrak{N}$ \rev{instead}, corresponding to ``gauging" the duality symmetry.

According to \cite{Antinucci:2023ezl}, it is also quite simple to describe the topological operators confined on $\calL_{\mathbf{B}, \psi}$. Let us consider only the case of a Fiber Functor. On the boundary condition $\calL_{\mathbf{B}, \psi}^{D}$, where $D$ stands for the Dirichlet boundary condition for $\mathfrak{a}$, the symmetry is a split 2-Group with action $\sigma$ and mixed 't Hooft anomaly:
\be
I = 2 \pi i \int A \cup_\sigma \fP_\psi(B) \, ,
\ee
where $\fP_\psi$ is the quadratic refinement of the symmetric form $\psi(B)(B')$.  Gauging $A$ to reach the Fiber Functor description gives a "non-invertible" 3-Group, described by surface operators:
\be
W_{2,a} = U_a + U_{\sigma(a)} \, , 
\ee
and an invertible line operator:
\be
H_1 \, ,
\ee
which can emanate from pointlike intersections of $W_{2,a}$. The objects give charges for the one-form symmetry and the duality symmetry respectively.
The presence of a 3Group impacts the fusion structure of defect multiplet operators on a symmetric boundary condition.
The simplest example is the case of $\abA=\bZ_2$, $\phi=1$, the symmetric defect corresponding to the dyonic boundary condition generated by the $(1,1)$ anyon. In this case $\sigma$ is trivial and one simply finds the 3-Group:
\be
d S = \fP(B) \, .
\ee
Pictorially, on the four dimensional boundary two $W_2$ surfaces --charged under the one-form symmetry-- intersect at a point. From here a duality charge $H_1$ emanates. Projecting this picture on $\calM$ we find that the intersection between two charged line operators $\phi_1$ at the boundary of $W_2$ carries non-trivial duality charge, see Figure \ref{fig:2Group}.
We now extend this logic to line and surface operators by performing the appropriate dimensional reductions.
\subsection{Line multiplets ($p=3$)}\label{ssec: lineduality} First let us classify line operators. The reduced DW theory is simply:
\be
S[S^2] = 2 \pi i \int b_0 \cup \delta c_2 + c_0 \cup \delta b_2 \, , 
\ee
Dirichlet boundary conditions correspond to specified holonomies:
\be
\exp(2 \pi i b_0) = \exp(2 \pi i \alpha) \, , \ \ \ \exp(2 \pi i c_0) = \exp(2 \pi i a) \, ,
\ee
these describe a dyon with charges $(a, \, \alpha)$. The canonical boundary condition $\calL_{sym}[S^2]$ is Dirichlet for both $c_2$ and $c_0$ and indeed an electric surface:
\be
U_a = \exp \left(2 \pi i a \int b \right) \, ,
\ee
can attach to a line giving it charge $a$. The dyon is duality invariant only if:
\be
- \phi^{-1}(\alpha) = a \, , \ \ \ \phi(a) = \alpha \, ,
\ee
that is: $2 a = 0$ and $\alpha = \phi(a)$. This admits solutions iff $\abA$ contains order two elements. This immediately shows that even an anomaly-free symmetry can forbid a (charged) invariant line. The first example is $\abA=\bZ_5$, with $\phi = 1$. In this case the condition $\sigma^2=-1$ boils down to $-1$ being a quadratic residue $\mod 5$, which has solutions $\sigma =2,3$. On the other hand, since 5 is an odd integer, the equation $2 a =0$ has only the trivial solution.

\subsection{Surface multiplets (p $=$ 2)}
Similarly we can study the dimensional reduction on $S^1$ of the DW theory:
\be
S[S^1] = 2\pi i \int b_1 \cup \delta c_2 + c_1 \cup \delta b_2 \, .
\ee
Denoting the surface and line operators by a quadruple $(a_1, \alpha_1; a_2; \alpha_2) \equiv (\mathbf{a}_1 ; \mathbf{a}_2)$ the braiding between a line and a surface is given by:
\be
\cB\left[(a_1, \, \alpha_1) , \, (a_2, \, \alpha_2) \right] = \alpha_1[a_2] \alpha_2^{-1}[a_1] \, .
\ee
Which defines an alternating pairing between $\abA \times \abA^{\vee} \equiv \mathbf{A}$ and its dual.

The duality symmetry acts on the quadruple $(a_1, \alpha_1; a_2, \alpha_2)$ by:
\bea
S_\phi : \ &(a_1, \alpha_1; a_2, \alpha_2) &&\longrightarrow (-\phi^{-1}(\alpha_1), \phi(a_1) ; -\phi^{-1}(\alpha_2), \phi(a_2) ) \, , \\
&(\mathbf{a}_1;\mathbf{a}_2) &&\longrightarrow (\Phi(\mathbf{a}_1);\Phi(\mathbf{a}_2)) \, .
\eea
and thus implements a 3d duality symmetry with the special choice $\chi_1=\chi_2=\chi$. A Lagrangian algebra $\calL_{\mathbf{B}, \psi}$\footnote{\rev{Here we will suppress the $S^1$ dependence to ease the notation a bit. All the algebras appearing are intended in the dimensionally reduced theory.}} is described by first choosing a subgroup $\mathbf{B} \subset \mathbf{A}$ of surface operators\footnote{Notice that this contains the algebras in the uncompactified theory by default.} and completing the spectrum with line operators in $N(\mathbf{B})$:
\be
\calL_{\mathbf{B}, \psi} = \left\{ (\mathbf{\beta}, \mathbf{b}) \, , \ \mathbf{b} \ \in \ \mathbf{B} , \, \ \mathbf{\beta} \ \in \ N(\mathbf{B})\right\} \, .
\ee
Furthermore we have a choice of fractionalization of surface junctions. At a three-valent junction labelled by $\mathbf{b_1}, \mathbf{b}_2$ we can insert a line operator $\psi(\mathbf{b_1}, \mathbf{b}_2) \in N(\mathbf{B})$. The invariant information in $\psi$ is contained in a fractionalization class $\psi \, \in \, H^2(\mathbf{B}, N(\mathbf{B}))$.\footnote{In particular we will need the case $\mathbf{B}=N(\mathbf{B})=\bZ_n$, in which case $\psi$ is just the Bockstein map:
\be
\psi(b_1,b_2) = \frac{b_1 + b_2 - [b_1 + b_2 \text{mod}(n)]}{n} \, .
\ee}

The symmetry in this case is always group-theoretical \cite{Decoppet:2023bay}, indeed the algebras:
\bea
&\calL_{0} &&= \left\{ (\mathbf{a}; \mathbf{0})  \, , \ \mathbf{a} \ \in \ \mathbf{A} \right\} \, , \ \ \ \text{and} \\
& \calL_{\mathbf{A}} &&= \left\{ (\mathbf{0};\mathbf{a}) \, , \ \mathbf{a} \ \in \ \mathbf{A}  \right\} \, , 
\eea
are always duality-invariant. They however break part of the symmetry generated by $b_1$ and $b_2$ respectively.

\vspace{2mm} \noindent \textbf{Duality-invariant algebras}
A generic duality-invariant algebra must satisfy:
\be
\Phi(\mathbf{B}) = \mathbf{B} \, , \ \ \ \Phi({N(\mathbf{B})}) = N(\mathbf{B}) \, ,
\ee
where the second condition follows automatically if the first is satisfied. Furthermore $\psi$ must transform covariantly under duality:
\be
\Phi^{-1}(\psi(\Phi(\mathbf{b}_1), \Phi(\mathbf{b}_2)) = \psi(\mathbf{b_1}, \mathbf{b}_2) \, .
\ee
Lastly, we describe the Fiber-Functor. We first need $\mathbf{B} = \left\{ (b, \theta(b) ) \, , b \in \abB \right\}$, with $\text{Rad}(\theta)=0$. 

To determine $N(\mathbf{B})$ notice that the braiding between an algebra surface and a generic line is:
\be
\alpha[b] \, \theta^{-1}(b)[a] \, ,
\ee
so $N(\mathbf{B})$ contains $\mathbf{B}$ in this case. Furthermore, if $N(\abB) \neq 0$, lines of the form $(0,\beta)$ are also present in $N(\mathbf{B})$, but this would contradict the assumption of a Fiber Functor. 
We conclude that $\abB=\abA$ and $N(\mathbf{B}) = \mathbf{B}$ since it saturates its dimension. Finally studying the braiding between lines and surfaces under duality we recover the conditions \eqref{eq: dualityinvfull}. In terms of $\theta$ we have $\sigma^{-1} \theta \sigma = \theta$.
This essentially gives back the dimensional reduction of the 4d Fiber Functors. After the dimensional reduction, however, there can be nontrivial duality-invariant classes $\psi$, so there are always \rev{at least as many} symmetric defects as boundary conditions. 

To conclude we notice that, for a duality-invariant algebra, further data might be needed in order to specify it completely. In \cite{Antinucci:2023ezl} these were described as an \emph{equivariantization} of $\calL$.\footnote{See also \cite{Cordova:2023bja} for a complementary perspective.} This describes a way in which the duality symmetry acts on the algebra data. In $3+1$ dimensions such characterization is incomplete, but it includes symmetry fractionalization classes. Importantly, in choosing such data, we must be sure that the $\bZ_4$ anomaly for the duality symmetry remains trivial. This greatly restricts the possible choices and we will not study it in detail in this work.

\vspace{2mm} \noindent \textbf{Example: SU(2)}\label{sssec: su2} Let us study concretely the example of $\abA=\bZ_2$, which is relevant for e.g. $SU(2)$ YM at $\tau=i$. The map $\phi =1$ is the identity one. $\mathbf{A} = \bZ_2 \times \bZ_2$ has five subgroups:
\be
\mathbf{B} = \left\{ \bZ_2 \times \bZ_2, \bZ_2^\abA, \bZ_2^{\abA^\vee}, \bZ_2^{D}, 0 \right\} \, .
\ee
The first and last entries describe the algebras $\calL_{\mathbf{0}}$ and $\calL_{\mathbf{A}}$, respectively which are duality-invariant. $\bZ_2^{\abA}$ is completed by an isomorphic $\bZ_2^{\abA}$ at the level of lines, into an algebra:
\be
\calL_{(\abA,0), \psi} = \left\{ (a,0;,a',0) \, , \ \ a, a' \ \in \ \abA \right\} \, .
\ee
There are two fractionalization classes, which are both duality-invariant:
\be
\psi_0(a_1,a_2) = 0 \, , \ \ \psi_1(a_1,a_2) = \frac{a_1 + a_2 - [a_1 + a_2 \, \text{mod}(2)]}{2}  \,\mod(2) \, . 
\ee
The algebras $\calL_{(\abA,0) , \psi}$ are mapped by duality to $\calL_{(0,\abA^\vee), \psi}$. Finally we study the dyonic algebra $\calL_{\mathbf{D}, \psi}$, generated by the dyon $(1,1)$, for which $N(\mathbf{D})=\mathbf{D}$. Again we have two duality invariant fractionalization classes. So we have \emph{two} Fiber-Functors:
\be
\calL_{\mathbf{D},0} \, , \ \ \ \calL_{\mathbf{D}, 1} \, .
\ee
We summarize the duality action on the 8 boundary conditions in the following diagram:
\bea
\begin{tikzcd}
    \calL_{\mathbf{0}} \arrow[loop left] & & \calL_{(\abA,0), 0} \arrow[leftrightarrow]{d}  & & \calL_{(\abA,0), 1}  \arrow[leftrightarrow]{d}  \\
    & & \calL_{(0,\abA^\vee), 0} & & \calL_{(0,\abA^\vee), 1} \\
    \calL_{\mathbf{A}} \arrow[loop left] & & \calL_{\mathbf{D}, 0} \arrow[loop left] & & \calL_{\mathbf{D},1} \arrow[loop right]
\end{tikzcd}
\eea
and the pattern of broken symmetry in Table \ref{tab: broken symm}. Notice that the presence of a duality-invariant class $\psi$ is very special to $\bZ_2$ among cyclic groups. Once can show by inspection that, for higher $n$, the class $\ell \beta$ is mapped to $- k \ell \beta$, with $k^2=-1 \, \text{mod}(n)$.
\begin{table}
    \centering
    $$
    \begin{array}{|c||c|c|c|c|c|c|>{\columncolor{yellow}}c|>{\columncolor{yellow}}c|} \hline \rule[-1em]{0pt}{2.8em} 
      \text{Algebra}   & \calL_{\mathbf{0}} & \calL_{\mathbf{A}} & \calL_{(\abA,0), 0} & \calL_{(\abA,0), 1} & \calL_{(0,\abA^\vee), 0} & \calL_{(0,\abA^\vee), 1} & \calL_{\mathbf{D}, 0} & \calL_{\mathbf{D}, 1} \cr  \hline \hline \rule[-1em]{0pt}{2.8em} 
       \bZ_2^{(0)} & \text{\XSolidBrush} & \text{\Checkmark}  & \text{\Checkmark} & \text{\Checkmark} & \text{\XSolidBrush} & \text{\XSolidBrush}  & \text{\Checkmark} & \text{\Checkmark} \cr \hline \rule[-1em]{0pt}{2.8em} 
      \bZ_2^{(1)} & \text{\Checkmark} & \text{\XSolidBrush} & \text{\Checkmark} & \text{\Checkmark} & \text{\XSolidBrush} & \text{\XSolidBrush} & \text{\Checkmark} & \text{\Checkmark}   \cr \hline \rule[-1em]{0pt}{2.8em} 
        \text{Duality} & \text{\Checkmark}  & \text{\Checkmark} & \text{\XSolidBrush} & \text{\XSolidBrush} & \text{\XSolidBrush} & \text{\XSolidBrush} & \text{\Checkmark} & \text{\Checkmark}   \cr \hline \rule[-1em]{0pt}{2.8em} 
        \text{\# Defect Vacua} & 2 & 1 & 2 &2 & 4 & 4 & 1 & 1 \cr \hline
    \end{array}
    $$
    \caption{Symmetries preserved by defect multiplets. Fiber Functors (symmetric defects) are in yellow. We also give the number of vacua on the defect.}
    \label{tab: broken symm}
\end{table}
For symmetric defects, after gauging the duality symmetry in the bulk, we can impose Neumann boundary conditions on the gauge field $\mathfrak{a}$, which then describes a local duality-charged operator on the defect. Upon this counting we find 10 surface defect multiplets for the $SU(2)$ duality symmetry:
\bea
\calL_{\mathbf{0}}^{D/N} \, , \ \ \ \calL_{\mathbf{A}}^{D/N} \, , \ \ \ \left(\calL_{(\abA,0),0/1} \oplus \calL_{(0, \abA^\vee), 0/1}\right) \, , \ \ \ \calL_{\mathbf{D}, 0/1}^{D/N} \, ,
\eea
the symmetric defects being $\calL_{\mathbf{D}, 0/1}^{N}$. 

Our results admits an interpretation in terms of 3d TFTs. Let us denote the background gauge fields for 0- and 1-form symmetries by $A$ and $B$ respectively. Up to unimportant normalization factors we have the following map between algebras and 3d partition functions:\footnote{The difference between the choice of fractionalization classes at the level of partition functions is immaterial on orientable manifolds, we write down the $A^3$ expression to orient the reader.}
\bea
&\calL_{\mathbf{0}} \longleftrightarrow \delta(A)  &&\calL_{\mathbf{A}} \longleftrightarrow \delta(B) \\ \rule[-1em]{0pt}{2.8em} 
&\calL_{(\abA,0),0} \longleftrightarrow 1  &&\calL_{(\abA,0),1} \longleftrightarrow \exp(2 \pi i \int A^3) \\ \rule[-1em]{0pt}{2.8em} 
&\calL_{(0, \abA^\vee),0} \longleftrightarrow \delta(A) \delta(B)  &&\calL_{(0, \abA^\vee),1} \longleftrightarrow \delta(A) \delta(B)  \\ \rule[-1em]{0pt}{2.8em} 
&\calL_{\mathbf{D},0} \longleftrightarrow \exp\left( 2 \pi i \int A B \right) &&\calL_{\mathbf{D}, 1} \longleftrightarrow \exp\left( 2 \pi i \int \left( A  B + A^3\right) \right) \, .
\eea
This classification has obvious applications to the study of e.g. Gukov-Witten operators in $\cN=4$ SYM at $\tau=i$ and we hope to report on this soon \cite{ACGRinprog}.

Let us also highlight some differences with respect to the top dimensional case i.e. boundary conditions. For the $SU(2)$ theory (on spin manifolds) there are three types of boundary conditions / TFTs \cite{Apte:2022xtu,Damia:2023ses}, described by 
\be
\calL_{0} \, , \ \ \calL_{\bZ_2} \, , \ \ \calL_{\bZ_2^D} \, .
\ee
corresponding to $\bZ_2^{(1)}$ gauge theory and the spin TFTs $\exp\left( \frac{2 \pi i s}{2} \int \fP(B) \right)$, $s=0,1$ for $\calL_{\bZ_2}$ and $\calL_{\bZ_2^D}$, respectively. The first two TFTs are exchanged under duality, while the third is the Fiber Functor. Including also the action of one-form symmetry, the allowed representations are a triplet and a singlet. 

The story for GW operators is different: for example we can either have doublets under the $\bZ_2$ symmetry (such as $\calL_{\mathbf{0}}$ and $\calL_{\mathbf{A}}$) or doublet under the duality symmetry ($\calL_{(\abA,0),0}$ and $\calL_{(\abA,0),1}$). Notice that even if some defects are duality-invariant, the defect $\cN$ can act nontrivially on the vacua. Consider $\calL_{\mathbf{0}}$, which has two defect vacua $|\pm\rangle$. Consistency implies that:
\be
\cN |\pm \rangle = |+\rangle + |-\rangle \, .
\ee
Importantly, such representation is forbidden for a purely two dimensional duality action \cite{Chang:2018iay}.
Focusing on GW operators in $SU(2)$ gauge theory \cite{Gukov:2006jk,Gukov:2008sn} the SSB of the zero-form symmetry implies that, in an electric description, the $\sigma$ model only couples to $SO(3) \subset SU(2)$. Parallel fusion with the bulk one-form symmetry defects gives rise to a new GW operator.

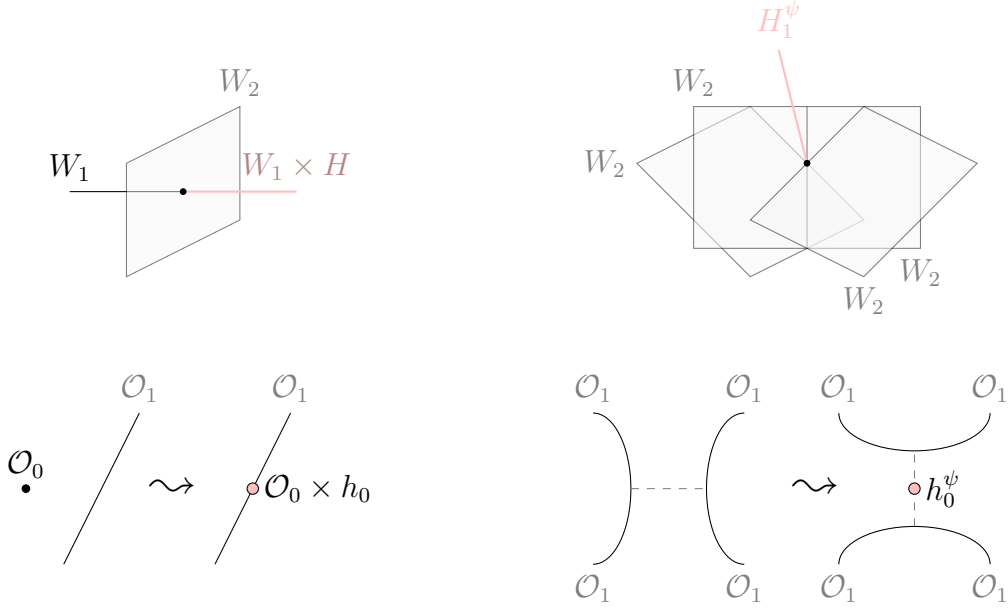
\begin{figure}
    \centering
    \begin{tikzpicture}[scale=0.75]
             \draw (-1,1.5) node[above] {$W_1$} -- (1,1.5);
        \draw[fill=white!97!black,opacity=0.5] (0,0) -- (2,1) -- (2,3) node[above,black] {$W_2$}-- (0,2)  --cycle;
         \draw[color=pink, thick] (1,1.5)  -- (3,1.5) node[above,black!30!pink] {$W_1 \times H$};
        \draw[fill=black] (1,1.5) circle (0.05);
   \begin{scope}[shift={(10,1)}]
      \draw[fill=white!97!black,opacity=0.5] (3,0)  -- (1,2)  -- (-1,1) node[left] {$W_2$} -- (1,-1) -- cycle;
        \draw[fill=white!97!black,opacity=0.5] (0,-0.5) -- (0,2) node[above] {$W_2$} -- (4,2) -- (4,-0.5) node[below] {$W_2$} --cycle;
        \draw[opacity=0.5] (2,-0.5) -- (2,2);
         \draw[fill=white!97!black,opacity=0.5] (1,0) -- (3,2) -- (5,1) -- (3,-1) node[below] {$W_2$} -- cycle;
         \draw[color=pink, thick] (2,1) -- (1.5,3) node[above] {$H_1^\psi$} ;
    \draw[fill=black] (2,1) circle (0.05);
   \end{scope}
    \end{tikzpicture}
    \\[0.5cm]
    \begin{tikzpicture}
    \draw (0,0) -- (1,2) node[above,opacity=0.5] {$\cO_1$};
    \draw[fill=black] (-0.5,1) node[above] {$\cO_0$} circle (0.05);
    \node[right] at (1,1) {\Large$\leadsto$};
    \begin{scope}[shift={(2,0)}]
      \draw (0,0) -- (1,2) node[above,opacity=0.5] {$\cO_1$};
      \draw[fill=pink] (0.5,1) node[right] {$\cO_0 \times h_0$} circle (0.075);
    \end{scope}

    \begin{scope}[shift={(7,0)}]
        \draw (0,0) arc (-90:90:0.5 and 1);
         \draw (2,0) arc (-90:-270:0.5 and 1);
         \node[opacity=0.5,below] at (0,0) {$\cO_1$};  \node[opacity=0.5,below] at (2,0) {$\cO_1$};  \node[opacity=0.5,above] at (0,2) {$\cO_1$};   \node[opacity=0.5,above] at (2,2) {$\cO_1$}; 
         \draw[gray, dashed] (0.5,1) -- (1.5,1);
         \node[right] at (2.5,1) {\Large$\leadsto$};
         \begin{scope}[shift={(3.25,0)}]
          \draw (0,0) arc (180:0:1 and 0.5);
         \draw (0,2) arc (180:360:1 and 0.5);    
          \draw[gray,dashed] (1,0.5) -- (1,1.5);
           \draw[fill=pink] (1,1) node[right] {$h_0^\psi$} circle (0.075);
           \node[opacity=0.5,below] at (0,0) {$\cO_1$};  \node[opacity=0.5,below] at (2,0) {$\cO_1$};  \node[opacity=0.5,above] at (0,2) {$\cO_1$};   \node[opacity=0.5,above] at (2,2) {$\cO_1$}; 
         \end{scope}
    \end{scope}
    \end{tikzpicture}
    \caption{Above, interpretation of the 2Group stucture on the symmetric boundary condition. Below, its projection onto the physical surface defect.}
    \label{fig:2Group}
\end{figure}

\paragraph{The defect multiplet structure} Let us comment on the multiplet structure on the two symmetric defects and how to distinguish them. The group theoretical-symmetry confined on $\calL_{\mathbf{D}}^D$ is by $\left(\bZ_2^{(0)} \times \bZ_2^{(1)}\right) \times \bZ_2^S$, with an anomaly:
\be
I = 2 \pi i \int A \cup \left( B_1 B_2 + \psi B_1^3\right) \, , \ \ \ \psi = 0,1 \, .
\ee
This can be derived from an $S^1$ reduction of (4.1) \cite{Antinucci:2023ezl} with minor changes.
If we make the $\bZ_2$ gauge field dynamical $A$ we will describe the set of topological operators on the Fiber-Functor boundary condition. There background 2-form $S$ for the dual symmetry satisfies \cite{Tachikawa:2017gyf}:
\be
dS = B_1 B_2 + \psi B_1^3 \, .
\ee
This is a special case of a 2Group. The topological operators describe the following charge multiplets:
\begin{itemize}
\item Topological defects $W_1$ and surface defects $W_2$ associated to $B_2$ and $B_1$ respectively, describe local $\cO_0$ and line $\cO_1$ local and line multiplets charged under the defect zero and one-form symmetry.
    \item Line defects $H_1$, associated to the background $S$, describe local defect multiplets $h_0$ charged under the duality symmetry.
\end{itemize}
The first term in the 2Group structure has the following interpretation: the cup product $B_1 B_2$ is activated once a surface $W_2$ intersects a line $W_1$ on the gapped boundary condition. From this intersection a line $H_1$ emanates. Projecting this onto the boundary shows that pushing $\cO_0$ onto $\cO_1$ decorates the local operator on the line with a duality charge.
The last term instead is a standard 2Group structure, describing how different resolutions of a 4-valent $\cO_1$ line junction on the defect leave behind a duality charge $\psi$. 
We summarize the two processes in Figure \ref{fig:2Group}.

\vspace{2mm} \noindent \textbf{Example: $SU(3)$} We also briefly outline an anomalous example, with $\abA=\bZ_3$. It is simple to see \cite{Damia:2023ses} that in this case we have no duality-invariant SPT, so we expect no symmetric GW defect to be allowed.

The story for the algebras $\calL_{\mathbf{0}}, \, \calL_{\mathbf{A}} , \, \calL_{(\abA, 0), \psi}$ and $\calL_{(0, \abA^\vee), \psi}$ is essentially the same. In this case we have $\psi_n = n \beta(a_1,a_2)$ with $n=0,1,2$ and $\beta$ the Bockstein map. The duality action exchanges $\calL_{(\abA,0),\psi}$ with $\calL_{(\abA^\vee,0),\psi}$ and $\calL_{(0,\abA^{\vee}), \psi}$ with $\calL_{(\abA,0),\psi^{-1}}$, so that it squares to charge conjugation.

Importantly, there are now two diagonal subgroups, $\mathbf{D}_1$ and $\mathbf{D}_2$, generated by $(1,1)$ and $(1,2)$, respectively. This time they are exchanged under duality. 
We conclude that indeed there is no room for a symmetric GW operator in $SU(3)$ SYM at $\tau=i$.

\section{'t Hooft Anomalies and obstructions to symmetric defects}\label{sec: anom}
There has been some debate in recent years about the correct generalization of the concept of an 't Hooft anomaly for a non-invertible symmetry. While for an invertible symmetry $\Gamma$ an 't Hooft anomaly is both an obstruction to gauging $\Gamma$ and to flow to a trivially gapped $\Gamma$-symmetric phase (an SPT), in the non-invertible case the two concepts do not coincide. Indeed it turns out that the latter -- which in technical terms describes a Fiber Functor for the symmetry category -- is stronger than the former \cite{Choi:2023xjw}. The relevance of Fiber Functors was pointed out in \cite{Thorngren:2019iar}. 
Fiber Functors also describe $\cC$-symmetric boundary conditions described by (higher) $\cC$ module categories with a single simple object. 

Thus, it is possible to interpret the presence of an 't Hooft anomaly (i.e. the lack of a Fiber Functor) as an obstruction to define a $\cC$-symmetric boundary condition. This obstruction has been decribed in \cite{Jensen:2017eof} for continuous symmetries and in \cite{Thorngren:2020yht} for discrete ones. Similar arguments also work in the case of interfaces, but what about extended defects of higher codimension?

The ``magnetic" description introduced in the present paper gives a clear -- albeit formal -- answer:
\begin{center}
    \textit{A theory admits symmetric defects} $\calD$ \textit{of codimension} $p$ \textit{iff the reduced symmetry} $\cC[S^{p-1}]$ \textit{admits a Fiber Functor}.
\end{center}
For invertible symmetries $\Gamma$ with 't Hooft anomaly $\omega$ this implies that a symmetric defect $\calD$ can exist if the dimensionally reduced anomaly vanishes, i.e.
\be
\omega [S^{p-1}] = d \eta \, .
\ee
A simple application concerns $\bZ_n$ anomalies for zero-form symmetries, which trivialise upon dimensional reduction on any sphere $S^{p-1}$. \rev{This implies that, while $\bZ_n$-invariant boundary conditions are generically forbidden, $\bZ_n$-symmetric defects can (and will) exist.}
\soutn{Indeed while $\bZ_n$-invariant boundary conditions are forbidden, $\bZ_n$-invariant local operators are allowed.}

In Section \ref{sssec: 3donef} we have found a different example, in which the dimensionally reduced anomaly does not trivialize. In that case, it implied that line operators cannot remain invariant after fusion with one-form symmetry generators.

In Section \ref{sssec: su2} we have instead shown an example where the number of symmetric defects is higher than the one for codimension one boundary conditions. The further splitting is completely due to the dimensional reduction procedure.

Finally it is worth remarking that this reasoning may fail if the reduced SymTFT has local topological operators. In this case insisting on having a \emph{indecomposable} boundary condition often forbids the presence of a Fiber Functor, even if the top dimensional theory was anomaly free. This is because the Fiber Functor boundary condition, once dimensionally reduced, typically is not indecomposable \soutn{in the presence of local topological operators}. We have seen examples of this phenomenon in \ref{ssec: 3d2d} and \ref{ssec: lineduality}.

\section{Conclusions and Future Directions}
In this Note we have give an alternative characterization of the realization of (generalized/higher) charges for categorical symmetry by analyzing gapped boundary conditions in the dimensionally-reduced SymTFT. 
We have given various examples of the strengths of our approach, which is especially suitable if the SymTFT has can be given a Lagrangian description.
While this work was mostly intended as a proof-of-concept, several interesting open questions remain:
\begin{itemize}
    \item Clearly the description provided in this Note is far from complete. Especially for higher categories the full layered structure of higher representations should come into play at some point. We have seen some glimpse of it in \ref{sssec: su2}.
    \item The \emph{existence} of higher charges does not imply that they necessarily are realized in a given physical theory. Thus the study of dynamical examples is paramount. In such context, the symmetry action on a defect and its defect operators should imply constraints and identities for e.g. the defect index \cite{Gadde:2013dda}.
    \item In \cite{Bhardwaj:2023bbf} the authors have explained how characterize possible symmetry-preserving defect transitions in terms of algebra embeddings. This clearly extends to defect by dimensional reduction. 
    It would be interesting to apply this to physically relevant systems \cite{Cuomo:2021rkm,Cuomo:2022xgw,Raviv-Moshe:2023yvq,Aharony:2022ntz,Copetti:2023sya,Ciccone:2024guw} for recent studies of RG flows on defects and boundaries in different contexts.
    \item Similarly, the SymTFT is extremely useful in describing the nontrivial properties of charged massive scattering \cite{Copetti:2024rqj,Copetti:2024dcz} in $(1+1)$d, such as modifications to the crossing symmetry.\footnote{See also \cite{vanBeest:2023dbu,vanBeest:2023mbs,Brennan:2023tae} for applications of generalized symmetries to the Callan-Rubakov effect and \cite{Mehta:2022lgq} for a violation of crossing symmetry in (2+1)d Chern-Simons-matter theories.} The formalism outlined here gives a natural avenue to generalize these properties to more interesting higher-dimensional systems.
    \item In this Note we have only studied the sphere reduction on $S^{p-1}$, describing isolated defects. It is likely that reductions on manifolds with non-trivial topology can describe interesting configurations, such as defect junctions. 
    \item Another natural generalization concerns the study of systems where the UV symmetry does not act faithfully on the gapless IR degrees of freedom. The way in which the kernel of this map is realized can be nontrivial and gives rise to (intrinsically) gapless SPTs \cite{Verresen:2019igf,Thorngren:2020wet,Li:2022jbf}. Their SymTFT realization is known in 1+1 d\cite{Wen:2023otf,Huang:2023pyk,Bhardwaj:2023bbf} and also in 3+1d \cite{Antinucci:2024ltv}. A defect igSPT prevents the defect to be screened by defect RG without incurring in spontaneous symmetry breaking. 
    \item Recently a SymTFT description of continuous symmetries (including some non-invertible ones) has been developed \cite{Antinucci:2024zjp,Brennan:2024tlw}. Out methods can be adapted to describe e.g. the $p=2$ surface charges in QED.
\end{itemize}

\paragraph{Acknowledgements} I am grateful to Francesco Benini, Michele Del Zotto, Lorenzo Di Pietro, Shota Komatsu, Kantaro Ohmori and Yifan Wang for discussions. To Andrea Antinucci, Lakshya Bhardwaj, Giovanni Galati, Daniel Pajer, Giovanni Rizi and Sakura Schafer-Nameki for collaboration on related projects and to Sara Oviglia for chromatic advice on the Figures. I am supported by STFC grant ST/X000761/1.

\bibliographystyle{ytphys}
\baselineskip=0.25\baselineskip
\small
\bibliography{symbib}

\end{document}